\documentclass[prb,amsmath, twocolumn, floatfix,showpacs]{revtex4}

\usepackage{graphicx} \usepackage{latexsym} \usepackage{amssymb}

\usepackage{color} \definecolor{blue}{rgb}{0,0,1}
\definecolor{red}{rgb}{1,0,0} \definecolor{green}{rgb}{0,1,0}

\newcommand{\lo}{\langle} \newcommand{\rc}{\rangle}
\newcommand{\ud}{{\mathrm d}}
\begin{document}
\title{Efficiency analysis of reaction rate calculation methods using
analytical models I: The 2D sharp barrier}

\author{Titus S. van Erp} \affiliation{Centrum voor Oppervlaktechemie
en Katalyse, K.U. Leuven, Kasteelpark Arenberg 23, B-3001 Leuven,
Belgium}

\begin{abstract}
We analyze the efficiency of different methods for the calculation of
reaction rates in the case of a simple 2D analytical benchmark
system. Two classes of methods are considered: the first are based on
the free energy calculation along a reaction coordinate and the
calculation of the transmission coefficient, the second on the
sampling of dynamical pathways.  We give scaling rules for how this
efficiency depends on barrier height and width, and we hand out simple
optimization rules for the method-specific parameters.  We show that
the path sampling methods, using the transition interface sampling technique, 
become exceedingly more efficient than the
others when the reaction coordinate is not the optimal one.
\end{abstract}

\maketitle
\section{introduction}
As Molecular dynamics (MD) is limited to microscopic systems and time
scales, most chemical or biological reactions can not be simulated
using straightforward MD.  One can literally wait ages before
detecting a single event in a typical computer simulation. In the
early 1930s, Wigner and Eyring made the first attempts to overcome
this problem by introducing the concept of the Transition state (TS)
and the so-called TS Theory (TST) approximation~\cite{E35,W38}.  Later
on, Keck \cite{Keck62} demonstrated how to calculate the dynamical
correction, the transmission coefficient. 
This work has later been extended by
Bennett~\cite{Bennet77}, Chandler~\cite{DC78} and
others~\cite{Yamamoto60,H38}, resulting in a two-step approach. First
the free energy as function of a reaction coordinate (RC) is
determined. This can be done by e.g. Umbrella Sampling (US)
\cite{TV74} or Thermodynamic Integration (TI)~\cite{CCH89}.  Then, the
maximum of this free energy profile defines the approximate TS
dividing surface and the transmission coefficient can be calculated by
releasing dynamical trajectories from the top.  This approach is, in
principle, exact and independent of the choice of RC.  However,
the method becomes inefficient when the transmission coefficient is
small.  A proper choice of the RC can maximize the transmission
coefficient and is hence crucial for the efficiency of the
method.

There exist different formalisms for the transmission coefficient
formula which differ in the way trajectories are counted. We discuss
the standard Bennett-Chandler (BC)~\cite{DC78}, 
the history dependent BC 
(BC2)~\cite{DC78}, and the effective positive flux (EPF)~\cite{Anderson95,ErpBol2004} formalism.
We show that the latter should
always be preferred due to a lower
average pathlength and a faster convergence.  However, whenever a lot
of correlated recrossings occur, the transmission coefficient will be
very low and all these methods become inefficient.  In
high dimensional complex systems it can be a very difficult task to
find a proper RC.  Moreover, whenever the dynamics is diffusive, even
an optimal RC can result in a very low transmission and hence a poor
efficiency.

A new approach came with 
Transition Path Sampling (TPS)~\cite{TPS98} that is not based on the
free energy barrier as starting point.  TPS is rather an importance
sampling of dynamical trajectories. Hence, it is a Monte Carlo (MC) 
sampling in path
space rather than phase space.  The TPS method has been advocated as a
method that does not need a RC and is akin to 'throwing ropes in the
dark'~\cite{Bolhuis02}. This might be true if one wants to sample a
set of reactive trajectories, but it is not  for the calculation of 
reaction rates.
In fact, the original approach to calculate reaction
rates within the framework of TPS required the definition of an order
parameter and the calculation of the reversible work when the endpoint
of the path is dragged along this parameter. For the sampling of
reactive pathways, the order parameter needs only to distinguish
between the two stable states. However, in the algorithmic procedure
to calculate reaction rates with TPS, the order parameter becomes very
similar to a RC.  Still, it has been speculated that this approach is
less sensitive to the problems related to an improper RC (or order
parameter). Indeed, in this article we prove for the first time that
this is true using the approach of Transition Interface Sampling
(TIS)~\cite{ErpMoBol2003}. TIS increases the efficiency of the
original TPS rate calculation considerably by allowing the pathlength
to vary and by counting only positive effective crossings.  The
overall reaction rate in TIS is obtained from an importance sampling
technique that uses a discrete set of interfaces between the stable
states.  Hence, TIS could be considered a dynamical analogue of US in
path space.  For diffusive systems the partial path TIS (PPTIS) was
invented that uses the assumption of memory loss~\cite{MoBolErp2004}.
In this article we discuss the case of sharp barriers.  Here
recrossings occur mainly due to the wrong choice of RC. In a follow-up
article we will treat the diffusive case.

Up to now, it is not clear how these methods compare in efficiency and
the need for benchmark systems has been put forward several times. It
is not always easy to perform comparative calculations since it is not
simple to know if each method is equally optimized for a specific
system. Therefore, in this paper we analyze a system for which the
efficiency of the methods can be calculated analytically with only a
few approximations.  This does not only give a transparent comparison
of the efficiency of the different methods, but also allows to obtain
scaling laws for how this efficiency changes as function of the
barrier height and width.  Moreover, we give some rules for how
the methods can be optimized, for instance, by choosing the proper
width of the US windows and the position of interfaces in TIS.  
These rules are important
for the simulation community, as they can be used as a rule of thumb
in daily practice, when any method needs to be optimized.

The principal component to measure the
efficiency of the methods will be the \emph{CPU efficiency time} 
$\tau_{\rm eff}$
which is the lowest computational cost needed to obtain an overall statistical
error equal to one. We give a detailed  analysis 
of how $\tau_{\rm eff}$ can be calculated for some very general cases
in the appendix sections~\ref{sgn} and ~\ref{secect}.
It also gives the important result of how
one should divide a total fixed simulation time over a set of
different simulations to obtain the best overall efficiency.  
In Sec.~\ref{secmet1}, we outline the first
class of methods, the reactive flux (RF) methods, and present their
principal formulas. In Sec.~\ref{secmet2}, we do the same for the
second class of methods, the path sampling methods.
Sec.~\ref{secsys} introduces the 2D benchmark system where
the angle $\theta$ indicates how far the chosen RC is deviated from 
the optimal one. 
Sec.~\ref{secres} is the main
section of our paper in which we apply the different methods of
Sec.~\ref{secmet1} and \ref{secmet2} to the analytical benchmark
system of Sec.~\ref{secsys}.  Finally, we discuss the important point of 
hysteresis
for the two types of methods and show that this is less likely to occur for the path sampling methods.
We summarize the results in
Sec.~\ref{seccon}.
Moreover, to support the readability of this paper we have added
a list of symbols and abbreviations in App.~\ref{alos} and \ref{aloa}. 

\section{First class of methods: Reactive Flux methods}
\label{secmet1}

\subsection{General formalism for RF methods}\label{secrf}
In all combined free energy and transmission coefficient based methods, 
the rate equation follows from $k =\tilde{k}(t')$
where the reaction rate $k$ is expressed as
a quasi-plateau value at a time $t'$ of a time dependent reaction rate
function $\tilde{k}(t)$. This function is given by the corrected flux through
an hypersurface $\{x | \lambda(x)=\lambda^* \}$, that is 
is a collection of phase points $x$,  
defined by the 
reaction coordinate $\lambda(x)$ and transition state (TS) value $\lambda^*$. The TS value
$\lambda^*$ is standardly taken as the maximum in the free energy profile along $\lambda$.
Both TST, BC, BC2 and EPF can be expressed as
\begin{align}
\tilde{k}(t)&= \frac{\lo \dot{\lambda}(x_0) \delta( \lambda(x_0)-\lambda^* )
\chi[X,t] \rc}{\lo \theta(\lambda^*-\lambda(x_0)) \rc},
\label{kgen}
\end{align}
where $\delta(\cdot)$ is the Dirac delta function, $\theta(\cdot)$ the
Heaviside step function, $x_t$ is the phase point at a time $t$, and $X$ is 
a  trajectory
that includes $x_0$. Ensemble averages $\lo \ldots \rc$ in phase and path space are defined 
in App.~\ref{subsensem}.
Eq.~(\ref{kgen}) measures the flux contributed by pathways leaving 
the surface $\lambda^*$ at $t=0$ under the influence of a correction functional
$\chi[X,t]$. 
The functional $\chi$ has different forms for TST,  
BC, BC2, and EPF.
The rate equation~(\ref{kgen}) can be rewritten as a product of two factors: 
the probability to be on top of the barrier times the 
transmission function $\tilde{\mathcal R}(t)$: 
\begin{align}
\tilde{k}(t) &= P_{A}(\lambda^*)  \tilde{\mathcal R}(t) 
\label{kpk}
\end{align}
with
\begin{align}
P_{A}(\lambda^*) &\equiv  \frac{\lo  \delta( \lambda(x_0)-\lambda^* ) \rc}{
\lo \theta(\lambda^*-\lambda(x_0)) \rc}.
\label{defRP}
\end{align}
and
\begin{align}
\tilde{\mathcal R}(t) &\equiv 
\lo \dot{\lambda}(x_0) \chi[X,t] \rc_{\delta(\lambda(x_0)-\lambda^*)}. 
\label{defR}
\end{align}
Here, $\lo \ldots \rc_{\delta(\lambda(x_0)-\lambda^*)}$  
implies that the ensemble $\{x_0\}$ is constrained at the surface 
$\{x_0| \lambda(x_0)=\lambda^* \}$. 
Substitution of Eqs.~(\ref{defR},\ref{defRP}) in Eq.~(\ref{kpk}) using Eq.~(\ref{wens}) gives back Eq.~(\ref{kgen}).

Eqs.~(\ref{kpk}-\ref{defRP}) show the two-step procedure. 
The probability $P_{A}(\lambda^*)$ and the time dependent
transmission function $\tilde{\mathcal R}(t)$ are calculated in two separate simulations.
As for the rate $k$, the unnormalized transmission coefficient ${\mathcal R}$ follows from a 
plateau in this time dependent function: ${\mathcal R}=\tilde{\mathcal R}(t')$. 
This factor corrects for the correlated recrossings.
In~\ref{subsfe}, we discuss the methods to compute  $P_{A}(\lambda^*)$ or, equivalently,
the free energy barrier $\Delta F =-k_B T \ln P_{A}(\lambda^*)$. Then, 
in~\ref{sectrans} we discuss the methods to determine the transmission coefficient ${\mathcal R}$.

\subsection{Free energy methods}
\label{subsfe}
\subsubsection{US using rectangular windows}
\label{susecUSr}
\begin{figure}[htbp]
\begin{center}
\includegraphics[width=6cm, angle=-0]{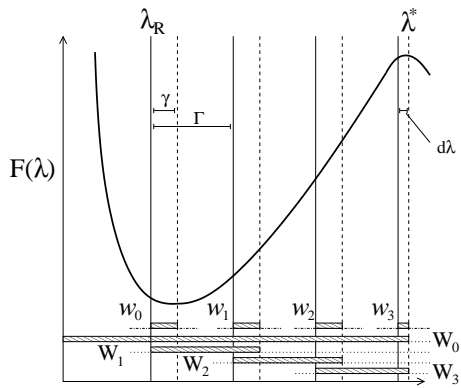}
\caption{Illustration of US using rectangular windows for the case $M=3$.
$\lambda_R$ is a surface in the reactant well, $\lambda^*$ is the TS. The
binary functions $w_i$ and $W_i$ are depicted below the free energy plot where
the gray areas indicate where $w_i, W_i$ equal unity.}
\label{umbrel}
\end{center}
\end{figure}

Define the following $2 M+2$ box functions:
\begin{align}
w_i(x)&=\theta[\lambda(x)- \lambda_R - i \Gamma] 
        \theta[\lambda_R+i\Gamma+\gamma-\lambda(x)], \nonumber \\
w_M(x)&=\theta[\lambda(x)+d\lambda-\lambda^*]
        \theta[\lambda^*-\lambda(x)],  \nonumber \\
W_0(x)&=\theta[\lambda^*-\lambda(x)], \label{windows1} \\
W_j(x)&= \theta[\lambda(x)-\lambda_R-(j-1) \Gamma]
\theta[\lambda_R+j \Gamma+\gamma  - \lambda(x)],
\nonumber 
\end{align}
with $0 \leq i < M$ and $0 < j \leq M$. Here $\lambda^*$ is the TS,  
$\lambda_R$ is a value in the reactant well and $d\lambda$ is a small
length scale. $\gamma$ and $\Gamma$ represent the dimensions of the 
US windows; $\Gamma+\gamma$ is the width of the window and $\gamma$ is the 
overlap such that $M \Gamma+d \lambda=\lambda^*-\lambda_R$ 
(See Fig.~\ref{umbrel}).
Neglecting higher orders terms in $d\lambda$, we can write
\begin{align}
P_{A}(\lambda^*)&=
\frac{1}{\ud\lambda}
\frac{\lo w_M \rc}{\lo \theta(\lambda^*-\lambda) \rc}
=\frac{1}{\ud\lambda} \lo w_M \rc_{W_0 }.
\label{FreeEnSam}
\end{align}
To calculate Eq.~(\ref{FreeEnSam}), we can simply run
an MD simulation and count the number of times that the transition state
region interval is visited. 
The weight function $W_0$ in the ensemble  acts like an infinite wall
at $\lambda=\lambda^*$ and prevents the unnecessary exploration of the
product region. However, as $\lo w_M \rc_{W_0 }$ is vanishingly small
for high barriers, this straight-forward method will usually fail. 

Using Eq.~(\ref{wens}) and the relations
$w_s W_s=w_s$, $w_{s-1} W_s=w_{s-1}$ for all $s$
we can rewrite Eq.~(\ref{FreeEnSam}) as
\begin{align}
P_{A}(\lambda^*)&=
\frac{1}{\ud\lambda}
\lo w_0 \rc_{ W_0 }
\prod_{s=1}^{M}
\frac{\lo w_s \rc_{ W_s }
}{
\lo w_{(s-1)} \rc_{ W_s }}.
\label{USrw}
\end{align}
The final property is now calculated from a series of simulations in which 
each pair $\lo w_s \rc_{ W_s}, \lo w_{s-1} \rc_{ W_s} \gg
\lo w_M \rc_{ W_0}$ so that they can be determined accurately.
The implementation of US using rectangular windows via MC is straightforward. 
The standard
MC sampling is performed starting from a point inside the window. As soon as the MC procedure generates a point outside this window, 
this point is automatically rejected and the old point 
is kept. If the procedure is performed by means of MD, the window boundaries simply act as infinitely 
hard walls. However, due to practical problems related to a discontinuous force profile, 
MD simulations are
usually performed with parabolic windows instead of rectangular ones.  

\subsubsection{US using a single biasing potential}
\label{susecUSb}
Instead of performing several simulations using rectangular windows, one 
can also use a single biasing function $\Omega(x)$:
\begin{align}
P_{A}(\lambda^*)&=  
\frac{1}{\ud\lambda}
\frac{\lo w_M \Omega^{-1} \rc_{\Omega W_0}}
{ \lo \Omega^{-1} \rc_{\Omega W_0}  }.
\label{USsb}
\end{align}
Again, the equivalence between Eq.~(\ref{FreeEnSam}) and Eq.~(\ref{USsb}) 
is easy to prove via 
Eq.~(\ref{wens}).
The advantage of Eq.~(\ref{USsb}) is
that the error does not propagate as in the case of several windows. On the other hand,
one needs  to have already a good idea about the shape of the barrier to construct a good bias $\Omega$.
The series of rectangular windows is a more robust way to explore the barrier region when no knowledge is
available.  The algorithmic procedure to sample points in the biased ensemble $\lo\ldots \rc_{\Omega W_0}$ is explained in Sec.~\ref{subsensem}.

\subsubsection{Other free energy methods}
\label{suseofree}
Many other methods exist for the calculation of free energy surfaces. 
TI is of equally importance and is based on the constrained sampling
at surfaces from which the free energy's derivative can be obtained
at a given value of $\lambda$. Integration of these derivatives
results in the requested free energy profile.
In addition, many variations of US have been devised, such
as
Wang-Landau sampling~\cite{WangLandau}, meta dynamics~\cite{LP02}, and flooding~\cite{flood},
where
the optimal biasing potential is created on the fly. 

\subsection{Transmission coefficient calculation\label{sectrans}}
\subsubsection{TST approximation\label{susecTST}}
In TST, all pathways leading towards the product site $B$ 
are assumed to stay in $B$ for a very
long time. 
The TST approximation uses Eq.~(\ref{defR}) with
\begin{align} 
\chi^{\rm TST}[X,t]=  \theta(\dot{\lambda}(x_0)).
\label{chiTST}
\end{align}
If the barrier is sharp and a proper RC is taken, 
TST is a very good approximation  
or can even be exact. 
The calculation of ${\mathcal R}$, in TST, requires a simple 
numerical or analytical calculation. For instance, 
suppose that $\lambda$ is a simple Cartesian
coordinate with mass $m$, then $\dot{\lambda}(x_0)=v$ and
\begin{align}
{\mathcal R}=\frac{\int_0^\infty {\mathrm d} v \, v e^{-\beta \frac{1}{2} m v^2} }{
\int_{-\infty}^\infty {\mathrm d}v \,  e^{-\beta \frac{1}{2} m v^2}}
=\frac{1/\beta m}{\sqrt{2 \pi/\beta m}} =\frac{1}{\sqrt{2 \pi \beta m}}.
\label{RTST}
\end{align}
Therefore, the free energy profile is sufficient
to obtain  $k$ whenever TST is valid.

\subsubsection{BC formalism\label{susecBC}}
The BC equation is obtained using
\begin{align}
\chi^{\rm BC}[X,t]
= \theta[\lambda(x_t)-\lambda^*] 
\label{chiBC}
\end{align}
in Eq.~(\ref{defR}).
The evaluation of $\tilde{\mathcal R}(t)$ in Eq.~(\ref{defR}) 
consists of releasing 
many trajectories that start from the top of the barrier. These trajectories
only make a contribution different from zero if they end up at the product side of the barrier.
It is important to note that a trajectory with $\dot{\lambda}(x_0) < 0$ that leaves the surface heading
to the reactant state $A$ but finally ends up at the
product state $B$ at a time $t$, gives a negative contribution. 
The final value, which has to be positive per definition, results from
a cancellation of positive and negative terms. 

\subsubsection{BC2 formalism\label{susecBC2}}
The BC2 equation uses 
\begin{align}
\chi^{\rm BC2}[X,t] &= \theta[\lambda(x_t)-\lambda^*] 
\theta[\lambda^*-\lambda(x_{-t})].
\label{chiBC2}
\end{align}
Here, besides ending in the product state, the trajectories integrated backward
in time also have to end in the reactant state to give a nonzero contribution
to Eq.~(\ref{defR}). However, here as well, the contribution
of some trajectories are negative. This happens when the systems starts 
with a negative $\dot{\lambda}(x_0)$, but the forward and backward trajectories end up in the product
$B$ and reactant
state $A$, respectively. These so-called S-curves (trajectories that cross
the TS surface more than 2 times within a short time) are 
less likely to occur for sharp barriers. 

\subsubsection{EPF formalism\label{susecEPF}}
\begin{figure}[htbp]
\begin{center}
\includegraphics[width=6cm, angle=-0]{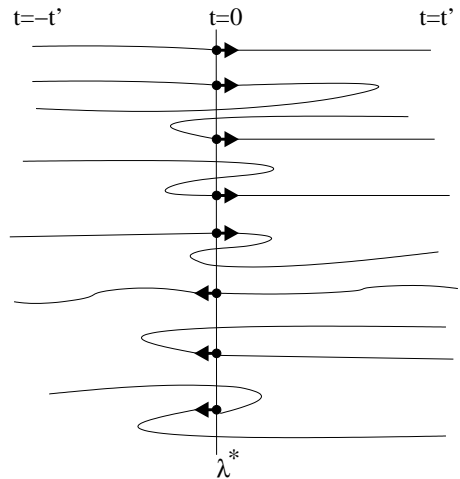}
\caption{Explanation of the transmission coefficient
method~(\ref{chiBC}-\ref{chiEPF}).
The dot on the surface $\lambda^*$ indicates the point of release at $t=0$.
The arrow indicates the direction of $\dot{\lambda}(x_0)$.
Trajectories are integrated forward up to $t=t'$ and backward in
time up to $t=-t'$. From top
to bottom, $\dot{\lambda}(x_0) \chi$ using BC Eq.~(\ref{chiBC}) results in
$(v,0, v,v,
v,0,-v,-v)$, the  BC2 Eq.~(\ref{chiBC2}) gives
$(v,0, 0,v,
v,0,0,-v)$, and the EPF expression Eq.~(\ref{chiEPF}) gives
$(v,0,0,0,
v,0,0,0$) with $v=|\dot{\lambda}(x_0)|$ the absolute 
velocity at the initial point.
}
\label{figepf}
\end{center}
\end{figure}

The EPF equation arises when we take $\chi[X,t]=\chi^{\rm EPF}[X,t]$ in Eq.~(\ref{defR}) with
\begin{align}
\chi^{\rm EPF}[X,t] &=
\begin{cases}
0, & \textrm{if } \dot{\lambda}(x_0) < 0,\\
 & \textrm{or if } \lambda_{t'} >  \lambda^* \textrm{ for any } t' \in [-t:0],\\
 & \textrm{or if } \lambda(x_t) < \lambda^*, \\
1, & \textrm{otherwise.}
\end{cases}
\label{chiEPF}
\end{align}
Despite the apparent mathematical more complicated form of Eq.~(\ref{chiEPF})
compared to Eqs.~(\ref{chiBC}) and (\ref{chiBC2}), the relation is quite is
natural. The approach counts only the first crossings
and only when they are in the positive direction
(See Fig.~\ref{figepf}). 
In the effective flux formalism (\ref{chiEPF}), all contributions are either 
zero or positive. 
In Ref.~[\onlinecite{ErpBol2004}], we presented a similar 
transmission coefficient formula. However, in this approach the pathway 
was stopped whenever the stable state regions $A$ and $B$ were reached. 
For a formal 
mathematical proof of the equivalence between BC- and EPF-type equations see 
Ref.~[\onlinecite{EricvE}].  

\subsubsection{Other transmission coefficient methods}
Some other relations for the transmission 
different from Eq.~(\ref{chiBC}-\ref{chiEPF}) have been proposed.
 For instance, Hummer~\cite{Hummer04} proposed a relation that uses all 
the crossing velocities, in case of correlated recrossings, 
 instead of just $\dot{\lambda}(x_0)$.
Ruiz-Montero \emph{et al.}~\cite{Ruiz} devised a transition zone method for 
diffusive barrier crossings. This will be treated in more detail in the follow 
up article that will discuss the diffusive barrier case.

\section{Second class of methods : path sampling methods \label{secmet2}}

\subsection{General formalism of TIS methods\label{susecgf}}
Suppose $\lambda_A$ and $\lambda_B> \lambda_A$ are such that any $x$ with
$\lambda(x) < \lambda_A$ is at the reactant side $A$ and 
any $x$ with $\lambda(x) > \lambda_B$  
is at the product side $B$ of the unknown 
optimal TS dividing surface. It is important to note that $\lambda(x)$
does not have to be a proper RC
to fulfill this criterion, but only has to distinguish between 
the two stable states.
The TIS expression is now given by
\begin{align}
\label{kabphiP}
k_{AB} &=\frac{\lo \phi_A \rc}{ \lo
h_{\mathcal A} \rc }  {\mathcal
P}_A(\lambda_{B}|\lambda_A).
\end{align}
Here 
$\phi_A$ gives the flux through interface $\lambda_A$ and
$h_{\mathcal A}$ is a history dependent function describing whether 
the system was more recently in $A$ or in $B$:
$h_{\mathcal A}=1$, for a phase point $x$  and its corresponding trajectory,
if the system was more recently in $A$ than in $B$ 
and $h_{\mathcal A}=0$ otherwise. 
In practice, the whole factor $\frac{\lo \phi_A \rc}{ \lo 
h_{\mathcal A} \rc }$  is calculated by
counting the 
number of crossings in a straight-forward MD simulation,
divided by the number of cycles with $h_{\mathcal A}=1$, divided by the 
time step $\Delta t$. The calculation of this  factor is very cheap as 
interface $\lambda_A$ 
is very close to the 
basin of attraction of state $A$.  On the other hand, the crossing probability  ${\mathcal P}_A(\lambda_{B}|\lambda_A)$
is a very small number and can not be accurately determined by straightforward MD.
This is the probability that whenever the system crosses interface $\lambda_A$, it will 
cross interface $\lambda_B$
before crossing interface $\lambda_A$ again. As interface $\lambda_B$ is an interface at 
the other side of the barrier, this probability is very small.
The TIS method overcomes this by defining $M-1$ interfaces  between $\lambda_A=\lambda_0$ and $\lambda_B=\lambda_M$.   
Then, the total crossing probability can be expressed in a formula 
that contains conditional short-distance
crossing probabilities~\cite{ErpMoBol2003}
\begin{align}
\label{TISEQ}
{\mathcal P}_A(\lambda_{B}|\lambda_A)= 
{\mathcal P}_A(\lambda_{M}|\lambda_0) &=
\prod_{s=0}^{M-1} {\mathcal P}_A(\lambda_{s+1}|\lambda_s).
\end{align}
Here, ${\mathcal P}_A(\lambda_{s+1}|\lambda_s)$ is a generalization of the previously described
overall crossing probability and denotes the conditional probability that, whenever the system
leaves state $A$ by crossing $\lambda_A$ and crosses $\lambda_s$ in turn, it will also 
cross $\lambda_{s+1}$ before returning to $A$.
If the distances $\lambda_{s+1}-\lambda_{s}$ are sufficiently small, the
probabilities ${\mathcal P}_A(\lambda_{s+1}|\lambda_s)$ will be large enough so that they can be computed
using a shooting algorithm~\cite{TPS98_2}.
The shooting algorithm takes a random time slice from the 
old existing path and makes a slight randomized modification of this phase point (usually only 
the momenta are changed). Then, this new phase point is used to propagate forward and backward in time
yielding a new trajectory. In TIS, this propagation of the trajectory is stopped whenever 
the system enters $A$ or $B$ or, equivalently, whenever $\lambda_0$ or $\lambda_M$ are crossed.
The pathway is  accepted only if the backward trajectory ends in $A$ 
\emph{and} the total 
trajectory  has at least one crossing with $\lambda_s$. The fraction of these 
paths that cross 
$\lambda_{s+1}$ as well, determines  ${\mathcal P}_A(\lambda_{s+1}|\lambda_s)$. 
Although the form of Eq.~(\ref{TISEQ}) deceptively resembles a 
Markovian factorization, no assumption has been made. As ${\mathcal P}_A(\lambda_{s+1}|\lambda_s)$
are not simple hopping probabilities, but incorporate the full history-dependence from 
$\lambda_A$ to $\lambda_s$, the relation is actually non-Markovian and exact~\cite{Moronithesis}.

\subsection{TIS biasing on $\lambda_{\max}$}
\label{biaslmax}
In analogy with US, instead of using a discrete set of interfaces, we could also bias the trajectory in a continuous way
using a bias on $\lambda_{\max}$~\cite{ErpMoBol2003}.
First we can write
\begin{align}
{\mathcal P}_A(\lambda_{M}|\lambda_0)= 
\lo \theta(\lambda_{\max}[X]-\lambda_B) \rc.
\label{Plm}
\end{align}
Here $\lambda_{\max}=\max\{ \lambda(x_t) | x_t \in X \}$ where the path $X$ is terminated 
when it leaves $A$ and enters region $A$ or $B$. Then, as in Eq.~(\ref{USsb}) we can write
\begin{align}
{\mathcal P}_A(\lambda_{M}|\lambda_0)= \frac{ \lo \theta(\lambda_{\max}[X]-\lambda_B)
\Omega^{-1}(\lambda_{\max}) \rc_{\Omega} }
{\lo \Omega^{-1}(\lambda_{\max}) \rc_{\Omega} }.
\end{align}
The algorithmic procedure would be the same as TIS without the interface crossing constraint.
Instead, the acceptance-rejection criterion is adjusted as explained in Sec.~\ref{subsensem}.
A continuous bias has been applied within the original TPS scheme.
In Ref.~[\onlinecite{Corcelli}] a bias on the end point of the path was used. Alternatively,
one can also bias the direction of the momenta change at the shooting point as was done in Ref.~[\onlinecite{MacFayden}].

\subsection{Other path sampling methods}\label{TISother}
The original TPS rate
calculation algorithm was introduced in 
Refs.~[\onlinecite{TPS98,TPS98_2,TPS99}]. It 
first creates ensemble of reactive trajectories of a fixed length. 
These trajectories should constitute a time interval that is longer than $t'$ 
where $\tilde{k}(t)$ reach its plateau.
Then, a second series of simulations is required that combines the MC of pathways with a US
technique. 
Also these simulations use a fixed pathlength, but these can be shorter 
using 
a rescaling approach~\cite{TPS99}. In the end, the final rate constant can be constructed
from the results of these two types of simulations.
Moroni showed that this algorithm is always less efficient than the TIS technique and that the 
computational advancement of TIS is at least a factor two~\cite{Moronithesis}.  
The TIS improvement is partly due to the flexible pathlength so that each individual path can be limited 
to its strictly necessary minimum. Moreover, TIS has abandoned the shifting moves
and has a stronger convergence (no cancellation positive and negative terms).
Depending on the system and the required accuracy, the TIS approach can 
easily become more than an order of magnitude 
faster than the original approach.

The PPTIS method reduces the average pathlength even 
further as compared to TIS using
the assumption of memory-loss~\cite{MoBolErp2004}. In this method, the trajectories do not have
to start at $\lambda_A$. Instead, an ensemble of trajectories is generated that start and end
at either $\lambda_{s+1}$ or $\lambda_{s-1}$, but must have at least one crossing with the middle
interface $\lambda_s$. From this, four crossing probabilities can be constructed that still inhibit some
history dependence. Once these are known for each $s$, the final overall crossing probability  
can be constructed via a recursive relation~\cite{MoBolErp2004}.
The Milestoning method~\cite{Milestoning} is
very similar to PPTIS, but relies on a stronger Markovian assumption
that the system remains in an equilibrium distribution at each
interface.
On the other hand, Milestoning uses time-dependent hopping probabilities
which supplies a very general way to coarse-grain a dynamical system.
The two aspects of PPTIS and Milestoning could be
combined as was suggested in Ref.~[\onlinecite{MoErpBol}].

Finally, we mention Forward Flux Sampling (FFS)~\cite{FFS,FFS2}.
FFS uses the same rate equation~(\ref{TISEQ}) as TIS, but the sampling
is different. In FFS~\cite{FFS,FFS2}, the crossing points at the next
interface of the 'successful pathways' are stored. The next
simulation uses this set of points to initiate new pathways. The
advantage is that FFS does not require any knowledge on the
distribution $\rho(x)$. This allows to treat non-equilibrium systems
as well. Moreover, FFS creates effectively more pathways than TIS
with the same amount of MD steps and does not rely on a Markovian
assumption as in PPTIS. However, the correlation between the
different pathways is much stronger than in TIS or PPTIS. Therefore,
FFS can only be applied when the process is
sufficiently stochastic.

\section{analytical 2D benchmark system}
\label{secsys}
\begin{figure}[htbp]
\begin{center}
\includegraphics[width=8cm, angle=0]{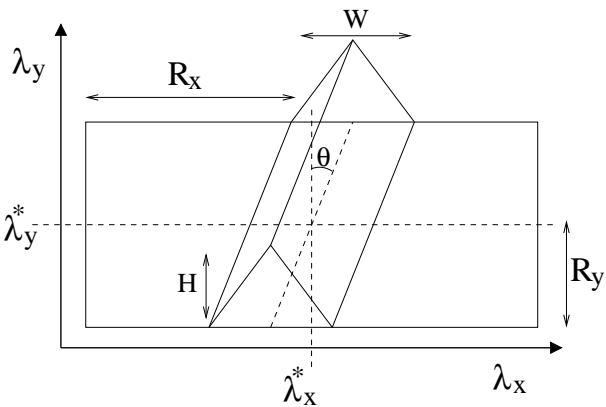}
\caption{The two-dimensional potential given by Eq.~(\ref{V2D}).
$H$ and $W$ are the barrier height and width.
$R_x$ and $R_y$ are the dimensions of the reactant (and product) well.
$\lambda_x$ is the assumed RC, while $\lambda_y$ is another important degree
of freedom. $\lambda_x^*$ and $\lambda_y^*$ are the maxima of the free energy
functions
for these coordinates.
 $\theta$ is
the angle between
the chosen reaction coordinate $\lambda_x$ and the optimal
one $\lambda_\perp(\lambda_x,\lambda_y)$ of Eq.~(\ref{lamortho}).
\label{f2d}}
\end{center}
\end{figure}

We consider the following 2D potential:
\begin{align}
V(\lambda_x,\lambda_y)&=
\begin{cases}
\infty, & \textrm{if } |\lambda_x-\lambda_x^*| > R_x, \\
V_{\rm bar}( \lambda_\perp) & \textrm{otherwise}\end{cases} \label{V2D}
\end{align}
with
\begin{align}
\lambda_\perp(\lambda_x,\lambda_y)&=(\lambda_x-\lambda_x^*)\cos \theta-
(\lambda_y-\lambda_y^*)\sin \theta \label{lamortho}
\end{align}
and
\begin{align}
V_{\rm bar}(\lambda)=
\begin{cases}
0, & \textrm{if } | \lambda| >   \frac{1}{2} W,\\
\big( 1- \frac{2 | \lambda|}{  W} \big) H  & \textrm{if }  | \lambda| <
\frac{1}{2} W. 
\end{cases}
\label{Vbar}
\end{align}
Here, $H$ is the height of our barrier, $W$ is the barrier width
and $R_x, R_y$ are the dimensions of the reactant region (See Fig.~\ref{f2d}). 
The chosen RC is $\lambda_x$, while $\lambda_y$ represents another important  
degrees of freedom.
$\lambda_\perp(\lambda_x,\lambda_y)$ is the unknown optimal RC as its direction 
is orthogonal to the barrier ridge, the exact TS dividing surface. 
The angle $\theta$ is, hence, a measure of the quality of the chosen RC.

To facilitate the analytics we assume a simplified dynamics:
The particles propagate without dissipation 
and  move only along the $\lambda_x$ direction.
Once they collide with the walls, they obtain a new random 
$\lambda_y$ coordinate
$\in [\lambda_y^*-R_y,\lambda_y^*+R_y]$
and velocities $\dot{\lambda}_x$
taken from a Maxwellian distribution. 
This type of dynamics satisfies ergodicity and ensures that the reaction rate 
is well defined for high barriers. 
Although the low dimensionality and  this dynamics might seem artificial,
this minimalistic model already illustrates clearly the problems
that occur in complex systems when no adequate RC can be found
as we will see in the forth-coming sections.  
Moreover, the potential can be viewed as a 
projection 
of a high-dimensional complex
system. 
In that case, Eq.~(\ref{V2D})
represents a free energy in which $\lambda_x$ and (the unknown but important
coordinate) $\lambda_y$ can be any non-linear function of all Cartesian 
coordinates. For instance, $\lambda_x$ could be the simple distance between
two atoms to describe the formation or breaking of a bond, while $\lambda_y$
represents a complex solvent rearrangement. 

The surface potential energy at the barrier $\lambda^*_x$ equals 
\begin{align}
V(\lambda^*_x,\lambda_y)= V_{\rm bar}( |\lambda_y-\lambda_y^*| \sin \theta  )= H(1-2
 |\lambda_y-\lambda_y^*| \sin \theta /W)
\end{align}
and, hence, 
\begin{align}
P_{A}(\lambda^*)&=\frac{1}{2 R_x R_y}
\int_{\lambda_y^*-R_y}^{\lambda_y^*+R_y} \ud \lambda_y  \, e^{- \beta V(\lambda_x^*,\lambda_y)} \nonumber \\=&
\frac{e^{-\beta H} \Big(e^{2 \beta H R_y \sin \theta /W } -1\Big)}
{2 \beta H R_x R_y \sin \theta /W}.
\end{align}
Here, we assumed that $R_x \gg W$.
The transmission coefficient can be obtained using Eq.~(\ref{defR}) and $\chi^{\rm BC2}=\chi^{\rm EPF}$
of Eqs.~(\ref{chiBC2},\ref{chiEPF}). The two equations are identical since
there are no trajectories that can cross the surface 
$\{x| \lambda_x(x)=\lambda_x^* \}$ more than two times before colliding with 
the outer walls.
Eqs.~(\ref{chiBC2},\ref{chiEPF}) have only non-zero contributions whenever
the backward and forward trajectory end, respectively, 
at the left  and right side of the 
barrier after a short
time $t'$, which is deterministically 
determined by $\dot{\lambda}_x(x_0)$. 
Hence,   
$\chi[X,t']=\chi(\dot{\lambda}_x, \lambda_y)|_{t=0}= 
\theta[\dot{\lambda}_x-\sqrt{2(H-V(\lambda_x^*,\lambda_y))/m}]$, which gives
\begin{align}
{\mathcal R}=\frac{ \int \ud \lambda_y \, {\mathcal R'}(\lambda_y) e^{-\beta V(\lambda_x^*,\lambda_y)}
}{
\int \ud \lambda_y \, e^{-\beta V(\lambda_x^*,\lambda_y)}
}
\label{R2D0}
\end{align}
with 
\begin{align}
{\mathcal R'}(\lambda_y)=\frac{\int_{\sqrt{2(H-V)/m}}^\infty \ud v \, v e^{-\beta \frac{1}{2} m v^2}
}{\int_{-\infty}^\infty \ud v \, e^{-\beta \frac{1}{2} m v^2}}
=
\frac{e^{- \beta( H - V(\lambda_x^*,\lambda_y)) }}{\sqrt{2 \pi \beta m}}.
\end{align}
Hence,
\begin{align}
{\mathcal R}&=
\frac{2 \beta H  R_y  \sin \theta/W}{\sqrt{2 \pi m \beta}} \Big( e^{2 \beta H  R_y \sin \theta /W} -
1\Big)^{-1} \label{R2D} 
\end{align}
which yields by Eq.~(\ref{kpk})
\begin{align}
k&=\frac{1}{\sqrt{2 \pi \beta m}}
\frac{ e^{-\beta H} }{  R_x }.
\label{k2D}
\end{align}
The reaction rate is, thus, independent of the angle $\theta$.

\section{Efficiency scaling}
\label{secres}
To quantify the efficiency of the different methods, we will calculate the
CPU efficiency time, $\tau_{\rm eff}$, that is defined 
as the minimal computational cost to
obtain an overall relative error equal to one.
The \emph{efficiency} is sometimes defined 
as the inverse $\tau_{\rm eff}^{-1}$
of this quantity~\cite{FFSeff,Mooij}.
In App.~\ref{secect}, we show how this quantity can be computed 
for some very generic cases. 
As the force calculations are the predominant steps
in any ab initio or large scale classical simulations,
all CPU values will be expressed
as an integer representing the number of required MD steps.
In this way, we obtain a measure that is 
independent of the computational resources.

In the following, we make two assumptions:
\begin{itemize}
\item The correlation number is assumed to be the same for each simulation $s$ 
in the simulation series: 
\begin{align} 
1+2 n_c^{aa}(s) =1+n_c^{ab}(s)+n_c^{ba}(s) \equiv  {\mathcal N_C}  
\textrm{ for each } s.
\label{assum} 
\end{align}
\item The mean cycle length $\tau_{\rm cyc}^{(s)}$ of the path-simulations and 
transmission coefficient calculations are proportional 
to the average pathlength $\tau_{\rm path}^{(s)}$ of the accepted paths: 
\begin{align} 
\tau_{\rm cyc}^{(s)} = 
\begin{cases}
1, & \textrm{ for MD or MC}, \\
\xi \tau_{\rm path}^{(s)} & \textrm{ for path sampling and transmission calc.}
\end{cases} 
\label{asstaucyc}
\end{align}
\end{itemize}
where $n_c^{aa},n_c^{ab},n_c^{ba}, \tau_{\rm cyc}$ are given in App.~\ref{sgn}
and \ref{secect} and $\tau_{\rm path}$ is the average number of MD steps
of the accepted trajectories.
Hence, we neglect the fact that ${\mathcal N_C}(s)$ and $\xi(s)$ can differ for each 
simulation $s$ in a simulation series. In fact, the equations that we will derive for TIS are true even if
a softer assumption is valid, i.e. that  ${\mathcal N_C} \times \xi$ 
is constant for each $s$.
In general, $\xi$ is smaller than 1 as rejected pathways are usually shorter 
than the accepted
ones. Some rejections are even immediate~\cite{ErpMoBol2003,ErpBol2004}
and do not require any  force calculations.
The transmission coefficient calculation has $\xi=1$ 
as each point on the TS, obtained from the first free energy calculation,
with randomized Maxwellian distributed velocities is
automatically accepted.
Still, the pathlengths $\tau_{\rm path}$ can differ. 

\subsection{RF methods: The free energy calculation for the 
$\theta=0$ case\label{eff1D}}
As explained in Sec.~\ref{secmet1}, the RF methods consist of two independent
types of calculations: the free energy and the transmission 
coefficient calculation. Moreover, there exist several techniques to 
determine these two quantities. 
As the TST approximation~(\ref{RTST}) is exact for the $\theta=0$ case
(which basically reduces the problem to one-dimension),
the only computational cost follows from the free energy calculation. 
Contrary, when $\theta$  is significantly larger than zero we can expect that
the transmission coefficient calculation is the dominant
computational factor even though the free energy calculation
becomes problematic as well (see Sec.~\ref{susechys}).  
Focusing on the most dominant contributions we will therefore assume 
$\theta=0$ for the free energy and $\theta >0$  for the transmission 
coefficient calculations.

\subsubsection{US using rectangular  windows}
\label{usrec}
First, to compare the enhanced efficiency of US techniques we need to know the
CPU efficiency time $\tau_{\rm eff}$ of straightforward MD.
Examination of Eq.~(\ref{FreeEnSam}) reveals that it simply
corresponds to the calculation of the ensemble average
of a binary function as in Eq.~(\ref{taubin}) with
$a=\lo w_M \rc_{W_0}$. Henceforth, 
using our assumptions~(\ref{assum},\ref{asstaucyc}) we have
$\tau_{\rm eff}= 
\Big( \frac{1-\lo w_M \rc_{W_0}}{\lo w_M \rc_{W_0}} \Big) 
{\mathcal N_C}
\approx \frac{\mathcal N_C}{\lo w_M \rc_{W_0}}
= {\mathcal N_C} R_x \,
e^{+\beta H}$. 
It is clear that the exponentially dependence on $\beta H$ make straightforward MD prohibitive 
for any high barrier
or low temperature system.

To obtain the overall efficiency for US using rectangular windows
as expressed in Eq.~(\ref{USrw}), we first need
the efficiency time $\tau_{\rm eff}^{(s)}$ for a single window. 
The principal result of this simulation $s$
equals 
$\lo w_s \rc_{W_s}/\lo w_{s-1} \rc_{W_s}$. The general approach
to calculate the efficiency time for 
a composite of two averages that are obtained simultaneously within
the same simulation is explained in Sec.~\ref{subscq}. 
In the App.~\ref{USCPU}, Eqs.~(\ref{covij}-\ref{partcov}), we derive 
that the efficiency time for this single window is given by
\begin{align}
\tau_{\rm eff}^{(s)}=
\begin{cases}
\frac{(1+e^{-\alpha \Gamma}) (e^{\alpha \Gamma}-e^{-\alpha \gamma})}{1-e^{-\alpha \gamma}} 
{\mathcal N_C}
&\textrm{if } \Gamma > \gamma, \\
\frac{(e^{\alpha \Gamma}-e^{-\alpha \gamma})  (1+e^{-\alpha \gamma}) (1-e^{-\alpha \Gamma}) }{(1-e^{-\alpha \gamma})^2} 
{\mathcal N_C}&\textrm{if } \Gamma < \gamma,
\end{cases}\label{taueffas}
\end{align}
where $\alpha=2 \beta H/W$. 
Here, $\alpha$ is a system-specific
parameter, ${\mathcal N_C}$ is an intrinsic of the simulation, and 
$\gamma$ and $\Gamma$ have to be 
optimized.
If we assume that $M$ is very large, we can neglect the difference of the first $W_0$ and last 
$w_M$ windows. 
Following Eq.~(\ref{effconst}), the overall 
efficiency time is given by $M^2 \tau_{\rm eff}^{(s)}$ and
as $M \approx W/(2 \Gamma)$ is not dependent on  $\gamma$, we can minimize $\tau_{\rm eff}^{(s)}$
as function of $\gamma$ to obtain also the lowest overall efficiency time. 
This optimum is achieved for half infinite windows $\gamma \rightarrow
\infty$  where
\begin{align}
\tau_{\rm eff}^{(s)}= (e^{\alpha \Gamma}-1) {\mathcal N_C}
\label{tinf}
\end{align}
for all $s$. 
As a result, the overall efficiency time  
equals 
\begin{align}
\tau_{\rm eff}= \Big(\frac{W}{2 \Gamma}\Big)^2(e^{\alpha \Gamma}-1) 
{\mathcal N_C}.
\label{tefftot}
\end{align}
Eq.~(\ref{tefftot}) has a minimum for
$\Gamma=1.6 \alpha^{-1}=0.8 \, W/\beta H$ which gives
\begin{align}
\tau_{\rm eff}&=  1.54 (\beta H)^2 {\mathcal N_C}.
\label{tauUSrw}
\end{align}
The efficiency time is quadratically dependent on the barrier height $H$ 
and the inverse temperature $\beta$. Compared to straightforward MD 
this is an enormous improvement. 
The optimal window boundaries imply that 
the fraction of phase points that is sampled in the right part of the window is given by 
$\lo w_s \rc_{W_s}=0.20$.  Hence, $\Gamma$ should be adjusted to have approximately 
one fifth of the sampled points in the most right up-hill part of the window.

\subsubsection{US using Single bias window} 
\label{subsubsbw}
In appendix~\ref{USCPU}, Eqs.~(\ref{sigmacovw},\ref{teffbcw}), we
derive the efficiency time for an ensemble average  $a=\lo \hat{a}(x) \rc$ 
when it is obtained via a different ensemble using 
a weight function $\Omega$: $a=\lo \hat{a} \Omega^{-1} \rc_\Omega/ \lo \Omega^{-1} \rc_\Omega$. Then, $\tau_{\rm eff}$ is given by:
\begin{align}
\tau_{\rm eff}&=\Big\{
\Big(\frac{\lo \hat{a}^2 \Omega^{-1}\rc \lo \Omega \rc}{ a^2}-1 \Big)
\Big[ 1+2 n_c^{bb} \Big] \nonumber \\
&+ \Big(\lo  \Omega^{-1}\rc \lo \Omega \rc-1 \Big)
\Big[ 1+2 n_c^{cc} \Big] \nonumber \\
&- 2 \Big( \frac{\lo \hat{a} \Omega^{-1}\rc \lo \Omega \rc}{a }-1 \Big)
\Big[ 1 + n_c^{bc}
+ n_c^{cb} \Big]
\Big\} \tau_{\rm cyc} \label{teffbias}
\end{align}
with $\hat{b}=\hat{a}\Omega^{-1}$
and  $\hat{c}=\Omega^{-1}$.
Note that the correlation functions $n_c$ (and  $\tau_{\rm cyc}$ for path sampling) can depend in principle on $\Omega$ as well. 
As Eq.~(\ref{teffbias}) does not change whenever
$\Omega$ is multiplied by a single factor,  
we apply the normalization  $\lo \Omega \rc=1$. 
Assuming that Eqs.~(\ref{assum},\ref{asstaucyc}) 
are also valid  in the biased ensemble, we write
\begin{align}
\tau_{\rm eff}&=\Big\{
\frac{\lo \hat{a}^2 \Omega^{-1}\rc }{a^2} 
+ \lo  \Omega^{-1}\rc  - 2  \frac{\lo \hat{a} \Omega^{-1}\rc }{a } 
\Big\} {\mathcal N_C}. \label{teffbias2}
\end{align}
If we assume that ${\mathcal N_C}$ has only a weak dependency on $\Omega$,
we can minimize Eq.~(\ref{teffbias2}) by taking the functional derivative 
$\delta \tau_{\rm eff} /\delta \Omega=0$
which gives (See Eqs.~(\ref{taueffbiasL}-\ref{w2}))
\begin{align}
\Omega(x) \sim |1-\frac{\hat{a}}{a}| \label{optw}
\end{align}
as optimal weight function.

Coming back to Eq.~(\ref{USsb}) using $\hat{a}=w_M$, the optimal bias 
follows directly from Eq.~(\ref{optw}) and  is given by
\begin{align}
\Omega(\lambda) = \frac{1}{2} \times \begin{cases}
 \frac{1}{ P_{A}(\lambda^*) \ud \lambda }
 &\textrm{if } \lambda^*-\ud\lambda < \lambda < \lambda^*,  \\ 
 1  &\textrm{otherwise},
 \end{cases}
\label{bias}
\end{align}
where we used Eq.~(\ref{FreeEnSam}) and $\lo w_M \rc_{W_0} = {\mathrm d} 
\lambda P_{A}(\lambda^*) \ll 1$. 
Substitution of Eq.~(\ref{bias}) and $\hat{a}=\hat{a}^2=w_M$  
in Eq.~(\ref{teffbias2}) and using that 
$\lo w_M \Omega^{-1} \rc \approx
2[ \ud \lambda P_{A}(\lambda^*)]^2$ and $\lo  \Omega^{-1} \rc \approx 2$
gives
\begin{align}
\tau_{\rm eff} & \approx  4 {\mathcal N_C}.
 \label{effUSsw}
\end{align}
Hence, 
a
scaling behavior  independent of the  
barrier height or system size can be achieved. 
We have assumed here that ${\mathcal N_C}$ is independent of the biasing 
function $\Omega$ which is only true for certain types of MC
such as those in which each cycle can be generated really independently (
hence ${\mathcal N_C}=1$).  In general, MC sampling is a diffusive type of 
motion and the bias~(\ref{optw},\ref{bias}) should also aid  the 
exploration from the top to the reactant well and back. 
Therefore optimal bias function should result in a more or less uniform
sampling distribution, which is achieved when $\Omega =e^{+\beta V}$.
Taking this into account, the efficiency time is a bit larger
$\tau_{\rm eff} \propto R_x/\ud \lambda$, but still independent
of $\beta H$.

\subsection{RF methods: The transmission coefficient 
calculation for the $\theta > 0$ case\label{eff2D}}
For the reasons explained in Sec.~\ref{eff1D}, here we will
study the case $\theta > 0$ and, in particular, we assume that 
$\beta H \sin \theta \gg 1$. 
Substituting $\hat{a}[X]=\chi[X]$ in Eq.~(\ref{taueff}) and using 
Eqs.~(\ref{std},\ref{defR},\ref{R2D},\ref{assum},\ref{asstaucyc}), 
we can write a general formula 
for the CPU efficiency time $\tau_{\rm eff}$ for  the RF 
method:
\begin{align}
\tau_{\rm eff} & =
\frac{ \lo \dot{\lambda}_x(x_0)^2 \chi^2 \rc_{\delta(\lambda_x(x_0)
-\lambda_x^*)} -{\mathcal R}^2}{{\mathcal R}^2} 
{\mathcal N_C} \tau_{\rm cyc}\nonumber \\
 & \approx  \frac{2 \pi m \beta}{\alpha^2} e^{2 \alpha}
{\mathcal N_C}  \tau_{\rm path}
 \lo \dot{\lambda}_x(x_0)^2 \chi^2 \rc_{\delta(\lambda_x(x_0)-\lambda_x^*)} 
 \label{teffR2D}
 \end{align}
where we used $\alpha \equiv 2 \beta H R_y \sin \theta /W \gg 1$. We remind you 
that $\xi=1$ for the transmission algorithm as all generated phase points on the surface $\lambda_x^*$ are accepted and
used to generate trajectories. In \ref{BC2D} and \ref{BC22D}, we will use formula (\ref{teffR2D}) 
to calculate the efficiency of the BC, BC2 and EPF method.

\subsubsection{BC}
\label{BC2D}
In the BC algorithm the pathways are propagated only forward in time. Say
$\Delta t \tau_{\rm esc}$ is the longest time that the system takes to leave the barrier when
released somewhere at the surface $\lambda_x^*$. Hence, to have a guaranteed
plateau in the $\tilde{R}(t)$ and $\tilde{k}(t)$ functions we 
simply have to integrate $\tau_{\rm esc}$ timesteps
so that
$\tau_{\rm path}=\tau_{\rm esc}$.
The second unknown factor in Eq.~(\ref{teffR2D}) is 
$\lo \dot{\lambda}^2 \chi^2 \rc_{\delta(\lambda(x_0)-\lambda^*)}$.
For the simplified dynamics we are considering
Eq.~(\ref{chiBC}) can be reduced to 
\begin{align} 
\chi^{\rm BC}=
\begin{cases}
1 & \textrm{if }  \lambda_y > \lambda_y^*  \textrm{ and }  
v > \sqrt{2 \Delta E /m},\\
   & \textrm{if } \lambda_y < \lambda_y^*  \textrm{ and }  
v > -\sqrt{2 \Delta E /m},\\
 0 & \textrm{otherwise}  
\end{cases} \label{chibc2d}
\end{align}
with $v=\dot{\lambda}_x(x_0)$ and $\Delta E(\lambda_y)=H-V(\lambda_x^*,\lambda_y)$.
Following the same lines as in Eqs.~(\ref{R2D0}-\ref{R2D}) gives
\begin{align}
\lo \dot{\lambda}_x(x_0)^2 \chi^2 \rc_{\delta(\lambda_x(x_0)-\lambda_x^*)} 
=\frac{
\int_{\lambda_y^*}^{\lambda_y^*+R_y} \ud \lambda_y  \chi'(\lambda_y) e^{-\beta V(\lambda_x^*,\lambda_y)}
}{  2 \int_{\lambda_y^*}^{\lambda_y^*+R_y} \ud \lambda_y e^{-\beta V(\lambda_x^*,\lambda_y)} }
\label{v2chiBC}
\end{align}
with 
\begin{align}
\chi'(\lambda_y)&= \frac{\int \ud v \, v^2 e^{-\beta \frac{1}{2} m v^2} \Big[ \theta(v-\sqrt{
\frac{2 \Delta E}{m}})+\theta(v+\sqrt{\frac{2 \Delta E}{m}})
\Big]}{\int \ud v e^{-\beta \frac{1}{2} m v^2}}.
\label{FBC}
\end{align}
In Eqs.~(\ref{v2chiBC},\ref{FBC}) we made use of the mirror symmetry  
along $\lambda_y^*$. As the Gaussian integral~(\ref{FBC}) has a symmetry as well along the $v=0$ axis, we
can rewrite Eq.~(\ref{FBC}) as follows
\begin{align}
\chi'(\lambda_y)&= \frac{ \int \ud v \, v^2 e^{-\beta \frac{1}{2} m v^2}  }{\int \ud v \, e^{-\beta \frac{1}{2} m v^2}} 
=\frac{\sqrt{2 \pi/ \beta^3 m^3 }}{\sqrt{2 \pi/\beta m}}=\frac{1}{\beta m}.
\label{FBC2}
\end{align}
Substitution of Eq.~(\ref{FBC2}) in Eq.~(\ref{v2chiBC}) reduces 
Eq.~(\ref{teffR2D}) to 
\begin{align}
\tau_{\rm eff}=\frac{ \pi \tau_{\rm esc}{\mathcal N_C}}{\alpha^2} e^{2 \alpha} =
\frac{\pi  \tau_{\rm esc}{\mathcal N_C}}{(2 \beta H R_y \sin \theta/W)^2
} e^{4 \beta H R_y \sin \theta/W}.
\label{tBC2D}
\end{align}
For large barriers, the exponential dependence on 
$4 H \beta \sin \theta R_y/W$ makes the method already
prohibitive for relatively small  angles $\theta$.

\subsubsection{BC2 and EPF}
\label{BC22D}
In BC2 the trajectories have to be followed until both forward and backward trajectories 
are no longer on the barrier. In EPF, the generated point on the surface heading toward 
reactant state are accepted, but assigned zero without further integration.  
Points on the surface with a positive velocity are first integrated backward and, then, integrated forward in time. This gives the advantage that whenever the backward trajectory recrosses the surface
$\lambda_x^*$ within a short period, this trajectory's contribution is assigned zero as well and the 
forward trajectory can be omitted. Hence, we have
$\tau_{\rm path}^{\rm BC2}=2 \tau_{\rm esc}$ and 
$\tau_{\rm path}^{\rm EPF} \lesssim \tau_{\rm esc}$
 
Moreover, as S-curves are absent in this system the two path-functionals are 
identical: $\chi^{\rm BC}[X]=\chi^{\rm EPF}[X]$ for all $X$. 
A non-zero contribution of $\chi$
occurs only when both the backward trajectory ends in the reactant state 
and the forward trajectory ends in the product state. This implies 
that the velocity $v$ should be positive with  kinetic energy higher
than $\Delta E \equiv H-V(\lambda_x^*,\lambda_y)$. Hence, 
\begin{align}
\chi^{\rm BC2/EPF}=\theta(v-\sqrt{\frac{2 \Delta E}{m}}).
\end{align}
Therefore, we can write the same equations as (\ref{v2chiBC},\ref{FBC}) with
$2 \theta(v-\sqrt{\frac{2 \Delta E}{m}})$ instead of $\theta(v-\sqrt{\frac{2 \Delta E}{m}})+\theta(v+\sqrt{\frac{2 \Delta E}{m}})$ in Eq.~(\ref{FBC}), 
which gives
\begin{align}
\chi'(\lambda_y)= 
\frac{1}{\beta m}\Big[ 2 \sqrt{\frac{\beta \Delta E}{ \pi}} e^{-\beta \Delta E} +
{\rm erfc}[\sqrt{\beta \Delta E} \Big]
\end{align}
We can neglect the ${\rm erfc}$-term by invoking~Eq.~(\ref{erfc}) and omitting the terms of order
${\mathcal O}([\beta \Delta E]^{-1/2})$. Substitution of this in Eq.~(\ref{v2chiBC}) gives
\begin{align}
\lo \dot{\lambda}(x_0)^2 \chi^2 \rc_{\delta(\lambda(x_0)-\lambda^*)} =
\frac{
\frac{1}{m}
\frac{e^{-\beta H}}{\sqrt{\beta \pi}}
\int_{\lambda_y^*}^{\lambda_y^*+R_y} \ud \lambda_y
\sqrt{\Delta E(\lambda_y)}
}{   \int_{\lambda_y^*}^{\lambda_y^*+R_y} \ud \lambda_y e^{-\beta V(\lambda_x^*,\lambda_y)} } \nonumber \\
= \frac{ \frac{2 R_y}{3 \beta m}
e^{-\beta H} \sqrt{\frac{\alpha}{\pi} }}
{R_y e^{-\beta H} e^\alpha/\alpha }=
\frac{2 }{3 \beta m} e^{-\alpha} \sqrt{\frac{\alpha^3}{\pi}}
\label{v2cBC2}
\end{align}
and, hence, Eq.~(\ref{teffR2D}) yields 
\begin{align}
\tau_{\rm eff}^{\rm EPF} \lesssim \frac{1}{2} \tau_{\rm eff}^{\rm BC2}=
\frac{4 \tau_{\rm esc} {\mathcal N_C}}{3 } \sqrt{\frac{\pi}{\alpha}} e^{\alpha}.
\label{tBC2EPF}
\end{align}
Although, the efficiency of Eq.~(\ref{tBC2EPF}) has been quadratically improved compared to 
Eq.~(\ref{tBC2D}), the exponential dependence of $\alpha=2 \beta H R_y \sin \theta/W$
makes this method prohibitive as well when $\theta$ is significantly different 
from zero. 

\subsection{TIS: the $\theta=0$ case}
\label{pathsam1D} 
As for the RF methods, the TIS methods consist also of two types of 
simulations: the initial flux through $\lambda_0$ and the crossing probability.
However, contrary to the RF methods, in TIS we might expect that the
crossing probability is always the computational  bottleneck.
As $\lambda_0$ is placed in the low potential energy region at the foot
of the barrier, this flux is easy to compute for all values of $\theta$.
We will henceforth concentrate on the crossing probability for the two cases
$\theta=0$ and $\theta>0$.

\subsubsection{standard TIS}
\label{TIS1D}
Say $a^{(s)}={\mathcal P}_A(\lambda_{s+1}|\lambda_s)$,   
$\lambda_0=\lambda^*-W/2$, and $\lambda_M=\lambda^*$.
For the pathlength we write
$\tau_{\rm path}^{(s)} 
\approx G (\lambda_s-\lambda_0)^g$ where the constants $G$ and $g$ 
will be determined later on.
As $a^{(s)}$ is basically an average of a binary function,
that is 1 for the successful trajectories and 0 otherwise, we can 
invoke Eq.~(\ref{taubin}) and use Eqs.~(\ref{assum},\ref{asstaucyc}): 
\begin{align}
\tau_{\rm eff}^{(s)}=\frac{1-a^{(s)}}{a^{(s)}} {\mathcal N_C} \xi 
G (\lambda_s-\lambda_0)^g.
\label{teffTISas}
\end{align}
${\mathcal P}_A(\lambda_{s+1}|\lambda_s)$ is 
the flux through $\lambda_s$ that reaches $\lambda_{s+1}$ divided by the total flux through
$\lambda_s$~\cite{ErpBol2004} for trajectories that come from $\lambda_0$.
Hence,
\begin{align}
a^{(s)}={\mathcal P}_A(\lambda_{s+1}|\lambda_s)= 
\frac{\lo  \dot{\lambda}_x(x_0) \delta(\lambda_x(x_0)-\lambda_s)  
h^b_{0,s} h^{f}_{s+1,0}  \rc}{\lo \dot{\lambda}_x \delta(\lambda_x(x_0)-
\lambda_s)  h^b_{0,s}   \rc}
\end{align}
where $h^{b/f}_{i,j}$ equals 1  only if the backward (forward) trajectory crosses $i$ before
$j$~\cite{ErpBol2004}. Otherwise it is 0.
For the case $\lambda_s < \lambda_{s+1} < \lambda_x^*$, we can write
$h^b_{0,s}=\theta(\dot{\lambda}(x_0))$ and
$h^f_{s+1,0}=\theta(\dot{\lambda}(x_0)-\sqrt{2 \Delta E/m})$
with $\Delta E =2 H (\lambda_{s+1}-\lambda_{s})/W$ the difference in potential energy
of the two surfaces. 
Hence,
\begin{align}
{\mathcal P}_A(\lambda_{s+1}|\lambda_s)= 
\frac{\int_{\sqrt{2 \Delta E/m}}^{\infty} \ud v \, v e^{-\beta \frac{1}{2} m v^2} }
{\int_0^\infty \ud v \, v e^{-\beta \frac{1}{2} m v^2} }=e^{-\beta \Delta E}.
 \label{PATIS}
\end{align}
We take an equidistant interface separation such that $\lambda_{s+1}-\lambda_s =
\frac{W }{2 M} \equiv \Delta \lambda$ and
$\lambda_s-\lambda_0=s \Delta \lambda$. Moreover, $[a^{(s)}]^M={\mathcal P}_A(\lambda_{M}|\lambda_0)=e^{- \beta H}$ or,
equivalently, $M=\beta H/|\ln a^{(s)}|$. 
This allows to rewrite Eq.~(\ref{teffTISas})
as
 \begin{align}
\tau_{\rm eff}^{(s)}=\frac{1-a^{(s)}}{a^{(s)}} |\ln a^{(s)}|^g \Big( \frac{W}{2 \beta H} \Big)^g s^g  {\mathcal N_C} \xi G.
\label{teffTISas2}
\end{align}
Since we know the efficiency times for each simulation $s$, 
we can calculate the 
total efficiency time of the whole simulation series.
It is, however, important to note that $\tau_{\rm eff}^{(s)}$ is not 
a constant as in Eq.~(\ref{taueffas},\ref{tinf}), but depends on $s$.
This raises an additional question of how one should divide a given total
simulation time among the different simulations. 
A logical choice would be to use the same simulation time for each $s$
or to adjust the simulation times  
to obtain the same relative error in each part.
We denote the total efficiency times using these two strategies 
$\tau_{\rm eff}'$ and $\tau_{\rm eff}''$.
Surprisingly, for this case the two approaches are equally efficient
and given by
\begin{align}
\tau_{\rm eff}'&=\tau_{\rm eff}''=\frac{1-a^{(s)}}{a^{(s)}}
\Big( \frac{\beta H}{|\ln a^{(s)}|} \Big)^2
 \Big( \frac{W}{2} \Big)^g \frac{  {\mathcal N_C} \xi G}{g+1} 
\end{align}
where we used Eqs.~(\ref{effprod12},\ref{effprod}) 
and $\sum_s^M s^g \approx M^{g+1}/(g+1)$.
As we show in App.~\ref{subsser},
these two approaches do not yield the optimum efficiency,
This would be attained when we assign each part $s$ a simulation time  
proportional to $\propto \sqrt{\tau_{\rm eff}^{(s)}}$ which yields
\begin{align}
\tau_{\rm eff}&=\frac{1-a^{(s)}}{a^{(s)}}
\Big( \frac{\beta H}{|\ln a^{(s)}|} \Big)^2
 \Big( \frac{W}{2} \Big)^g \frac{ 4 {\mathcal N_C} \xi G}{(g+2)^2}.
\label{effTIStot}
\end{align}
Minimizing Eqs.~(\ref{effTIStot}) with respect to $a^{(s)}$ 
shows that $\tau_{\rm eff}$ reaches a minimum for $a^{(s)}\approx 0.2$. 
Hence, the TIS procedure is optimized  when approximately one out of five 
trajectories are successful.   See the correspondence with US 
sampling~\ref{usrec} 
where one fifth was also the optimum for fraction of sampling point 
in the right uphill part of the window.
Using $a^{(s)}=0.2$ in Eq.~(\ref{effTIStot}) gives
\begin{align}
\tau_{\rm eff}&= 6.18 
\Big(\frac{W}{2}\Big)^g \frac{ (\beta H)^2  }{(g+2)^{2}} G \xi {\mathcal N_C}
\label{effTIS}
\end{align}
and $\tau_{\rm eff}'=
\tau_{\rm eff}''=\frac{(g+2)^2}{4 (g+1)}\tau_{\rm eff}$.
In practice, we have found a linear dependence ($g=1$) 
of the pathlength on a steep barrier\cite{ErpMoBol2003} and quadratically 
dependence ($g=2$) on a diffusive barrier~\cite{MoBolErp2004}. Hence, 
$\tau_{\rm eff}$ is about 12 \% and 
33~\% lower than $\tau_{\rm eff}'$ and $\tau_{\rm eff}''$. 
Note that Eq.~(\ref{effTIS}) has the same quadratically dependence on 
$\beta H$ as US~(\ref{tauUSrw}). 
One should realize that 
normally  ${\mathcal N_C}^{\rm MD/MC} \gg  {\mathcal N_C}^{\rm TIS}$ and that 
${\mathcal N_C}^{\rm MD/MC}$ usually has a strong $W$ dependence except for 
exceptional cases where MC cycles can be generated really independently. 
Hence, the efficiency scaling of TIS is comparable with that of US sampling using rectangular windows.
As US has more flexibility to reduce
${\mathcal N_C}$ than TIS to reduce $\tau_{\rm path}$, US is probably a bit 
more efficient than TIS by a single prefactor.

In this specific system, the $H$ dependence of the TIS efficiency 
is even a bit favorable than $g=1$.
In the App.~\ref{app_effTIS}, Eqs.~(\ref{TISpath}-\ref{erfc})  we derive that
\begin{align}
G \approx \frac{2}{\Delta t} \sqrt{\frac{Wm}H}, \qquad g \approx \frac{1}{2}
\label{GgTIS}
\end{align}
yielding  
\begin{align}
\tau_{\rm eff}&= \frac{1.40}{\Delta t} W \sqrt{m} \beta^2 H^{\frac{3}{2}} \xi {\mathcal N_C}.
\label{teffTISgG}
\end{align}
Due to a decreasing average pathlength,
TIS seems to have a slightly better scaling as function of 
$H$ than US using rectangular windows~(\ref{tauUSrw}). 
Assuming a lower prefactor for US,
this will imply that the TIS efficiency approaches the US efficiency 
Eq.~(\ref{tauUSrw}) at 
increasing  barrier heights.
However, Eqs.~(\ref{GgTIS},\ref{teffTISgG}) 
break down for
very high barriers as the average 
TIS pathlength $\tau_{\rm path}$ can, of course, in practice 
never decrease below one MD step.

\subsubsection{TIS biasing on $\lambda_{\max}$}
We can exactly follow the same lines as 
Eqs.~(\ref{teffbias},\ref{effUSsw}) which gives for the ideal biasing function
\label{biasmax}
\begin{align}
\Omega(\lambda_{\max}) = \frac{1}{2} \times \begin{cases}
 \frac{1}{ {\mathcal P}_A (\lambda_B | \lambda_A) }
 &\textrm{if }   \lambda_{\max} > \lambda_B, \\ 
 1  &\textrm{otherwise}
 \end{cases}
\label{TISbias}
\end{align}
and the overall efficiency
\begin{align}
\tau_{\rm eff}=4 {\mathcal N_C} \xi \tau_{\rm path} \label{effTISsw}.
\end{align}
Hence, the ideal bias function (\ref{TISbias}) allows to obtain a  scaling 
independent of $\beta H$ as 
in Eq.~(\ref{effUSsw}).
Note once more that generally ${\mathcal N_C}^{\rm MD/MC} \gg
{\mathcal N_C}^{\rm TIS}$. 
As for US, if we take into account the diffusive behavior of the sampling, 
${\mathcal N_C}$ is likely very large when the bias~(\ref{TISbias}) is used.
This bias favors only trajectories that reach state $B$, but does not aid the system 
to climb the  barrier in successive cycles.
Therefore, a bias that generates a more uniform distribution is more advantageous than 
Eq.~(\ref{TISbias}). However, 
this does not affect the scaling dependency on $\beta H$.

\subsubsection{other path sampling methods}
The estimation of the optimal interface separation on a straight slope 
is a bit difficult
for path sampling methods like PPTIS, Milestoning, and FFS
(Note that the PPTIS memory-loss assumption is satisfied 
on the strictly increasing barrier even if the
system is not diffusive~\cite{MoBolErp2004}).
An efficiency analysis of FFS~\cite{FFSeff} using the 
approximation~(\ref{assum_unc}) revealed that $\tau_{\rm eff}$ is constantly 
decreasing as function of the number of interfaces. The same result would
be valid for PPTIS. However, as correctly stated in Ref.~[\onlinecite{FFSeff}],
the apparent conclusion, that the computational cost can be made vanishingly 
small using an infinite set of infinitesimal spaced interfaces, is purely
artifical. If one takes into account that there is a minimum path length
(of at least 1 MD step), also the PPTIS and FFS show a quadratic 
dependence on $\beta H$. Considering the lower path length, 
the efficiency is likely to be very close
to US using rectangular windows. 

\subsection{TIS: the $\theta>0$ case}\label{TIS2D}
\begin{figure}[htbp]
\begin{center}
\includegraphics[width=8cm, angle=0]{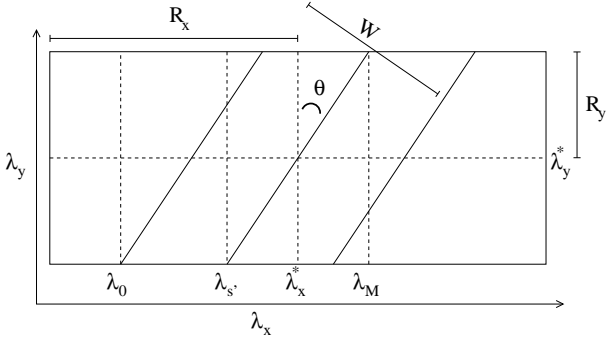}
\caption{Top view of the two-dimensional barrier and the
position of the interfaces
$\lambda_0, \lambda_{s'}, \lambda^*$, and $\lambda_M$.
}
\label{top2D}
\end{center}
\end{figure}

We take $\lambda_0=\lambda_x^*- W/(2 \cos \theta) - R_y \tan \theta$ and 
$\lambda_M=\lambda_x^*+R_y \tan \theta$ (See Fig.~\ref{top2D}). 
With these outer-interface $\lambda_x < \lambda_0$ 
ensures that the system is at the flat left side of the barrier and $\lambda_x> \lambda_M$ 
ensures that the system has crossed the barrier ridge. 
As in Eq.~(\ref{R2D}) we can write
\begin{align}
P_{\mathcal A}(\lambda_{s+1}|\lambda_s) =
\frac{\int \ud \lambda_y \,  P'_{\mathcal A}
(\lambda_{s+1}|\lambda_s; \lambda_y)  
e^{-\beta V(\lambda_s,\lambda_y)} }{
\int \ud \lambda_y \,   
e^{-\beta V(\lambda_s,\lambda_y)} } 
\end{align}
with $P'_{\mathcal A}(\lambda_{s+1}|\lambda_s; \lambda_y)$ the crossing probability along the line $\lambda_y$.
Now, the optimization of the full TIS algorithm would yield an extended and tedious calculation. 
Therefore, we derive an upper bound of the CPU efficiency time instead which is relatively easy. 
As in Eq.~(\ref{PATIS}), we can write 
for $P'_{\mathcal A}(\lambda_{s+1}|\lambda_s; \lambda_y)$:
\begin{align}
P'_{\mathcal A}(\lambda_{s+1}|\lambda_s; \lambda_y)={\min}[1,
e^{-\beta[V(\lambda_{s+1},\lambda_y)-V(\lambda_{s},\lambda_y)]}].
\end{align}
The additional ${\min}$-term compared to Eq.~(\ref{PATIS}) is because 
the potential energy can decrease 
from $\lambda_s$ to $\lambda_{s+1}$ along some 
coordinate $\lambda_y$.
Using equidistant interfaces with spacing $\Delta \lambda = (\lambda_M-\lambda_0)/M$,
we have $V(\lambda_{s+1},\lambda_y)-V(\lambda_{s},\lambda_y) \leq 
2 \Delta \lambda \cos \theta/ W$. 
Hence, $P'_{\mathcal A}(\lambda_{s+1}|\lambda_s; \lambda_y) \geq e^{-2 \beta \Delta \lambda H
\cos \theta/ W}$ and
\begin{eqnarray}
P_{\mathcal A}(\lambda_{s+1}|\lambda_s) > e^{-2 \Delta \lambda H \cos \theta/ W}
\label{Pg}
\end{eqnarray}
We remind you that the higher $a^{(s)}=P_{\mathcal A}(\lambda_{s+1}|\lambda_s)$
the lower $\tau_{\rm eff}^{(s)}$ due to Eq.~(\ref{taubin}). 
The equal sign in Eq.~(\ref{Pg}) for all 
$s$,  would correspond to the $\theta=0$ case  with barrier height 
$H'= 2 M \Delta \lambda H\cos \theta/ W =H(1+4 R_y \sin \theta/ W)$ and barrier width
$W'=W/ \cos \theta + 4 R_y \tan \theta$.
Therefore, we can simply invoke Eq.~(\ref{effTIS}) and write
\begin{align}
\tau_{\rm eff} & <  6.18 
\Big(\frac{W'}{2}\Big)^g 
 \frac{ (\beta H' )^2  }{(g+2)^{2}} G \xi {\mathcal N_C} \label{efftis2D}
\end{align}
which has only a quadratic dependence on $\beta H' \sim \beta H R_y \sin \theta/W$.  This is 
far superior to exponential scaling of Eqs.~(\ref{tBC2D},\ref{tBC2EPF}).   
As a result,  for high barriers and non-vanishing angles $\theta$,
the TIS method becomes orders of magnitude more efficient than the reactive flux methods. 

Of course, one might object that the reactive flux methods 
for this 2D system improve dramatically if we would 
chose
$\lambda_\perp$ instead of $\lambda_x$ as RC. 
The Reactive Flux efficiency is then again identical to the $\theta=0$ case.
However, this is exactly the crucial point.  It is quite simple to find a proper RC in a low dimensional 
system, but in high dimensional complex systems this is certainly not the case. 
Some methods have been devised that systematically search for RCs, but they have their limitations. 
The optimal hyperplane method~\cite{SMJ94,MJS95,HH01} and
the string method~\cite{ERVE02}
rely on harmonic approximations 
and on the 
assumptions that these hyperplanes can be described as a linear combination of Cartesian coordinates.
Complex reaction mechanism, notably chemical reactions in solution, require a highly nonlinear function as RC.  
The inclusion of important solvent degrees of freedom is not an easy task. Some success has been made using the coordination number as 
RC~\cite{Sprik98,Sprik2000}. However, the influence of the solvent can be 
more subtle. In Ref.~[\onlinecite{ErpMeij04}], 
we found that electric contributions due to nearby spontaneous formations of tetrahedral ordered water molecules 
can be crucial to give a last push over the potential energy barrier. 
To incorporate such an effect in a one-dimensional RC would be an enormous task and can not be
made without a priori insight in the mechanism. Automated multidimensional US sampling
approaches~\cite{LP02}  allow to explore the free energy surface  efficiently in a predetermined set of 
order parameters. From the reduced free energy potential the most optimal one-dimensional 
RC could be estimated~\cite{Ensing}. However, as the predetermined order-parameter space is still limited
to e.g. 6 coordinates, it is still likely that important coordinates such as solvent degrees of 
freedom can be missed. 

\section{Hysteresis}
\label{susechys}
Up to now, we have given expressions for  the efficiency times while
treating the effective correlation ${\mathcal N_C}$ as a simple constant factor.
The calculation of this factor is difficult as it 
depend on the intrinsic 
diffusive behavior of the MC/MD sampling itself. Fact is that ${\mathcal N_C}$
can be significant larger 1 and usually has a scaling dependence
on some system parameters (like $R_x, R_y$, $W$, and $\theta$). 
Especially, the 
$\theta$-dependence can be severe and will be discussed in this section.

US sampling and TI constrain the system in a small window or on a surface that drags the system
over the barrier. We have assumed that the time consuming step in the 
rate calculation for the
2D barrier is the transmission coefficient calculation. In fact, 
the calculation of the free energy
barrier can also be very hard due to an improper RC. 
This problem is known as the hysteresis problem
which basically results in a diverging  ${\mathcal N_C}$. Evidently, one could 
ask whether 
this problem occurs in TIS as well. If this would be the case, the TIS efficiency
would be much less advantageous as suggested by Eq.~(\ref{efftis2D}). 

We will show that sampling of paths instead of phase points alleviates or even eliminates
the hysteresis problem entirely. The hysteresis problem does not occur in  the 2D system 
we are considering, but can still persist, although still less than in free energy methods, for systems that have multiple reaction channels.
 
\begin{figure}[htbp]
\begin{center}
\includegraphics[width=6cm, angle=0]{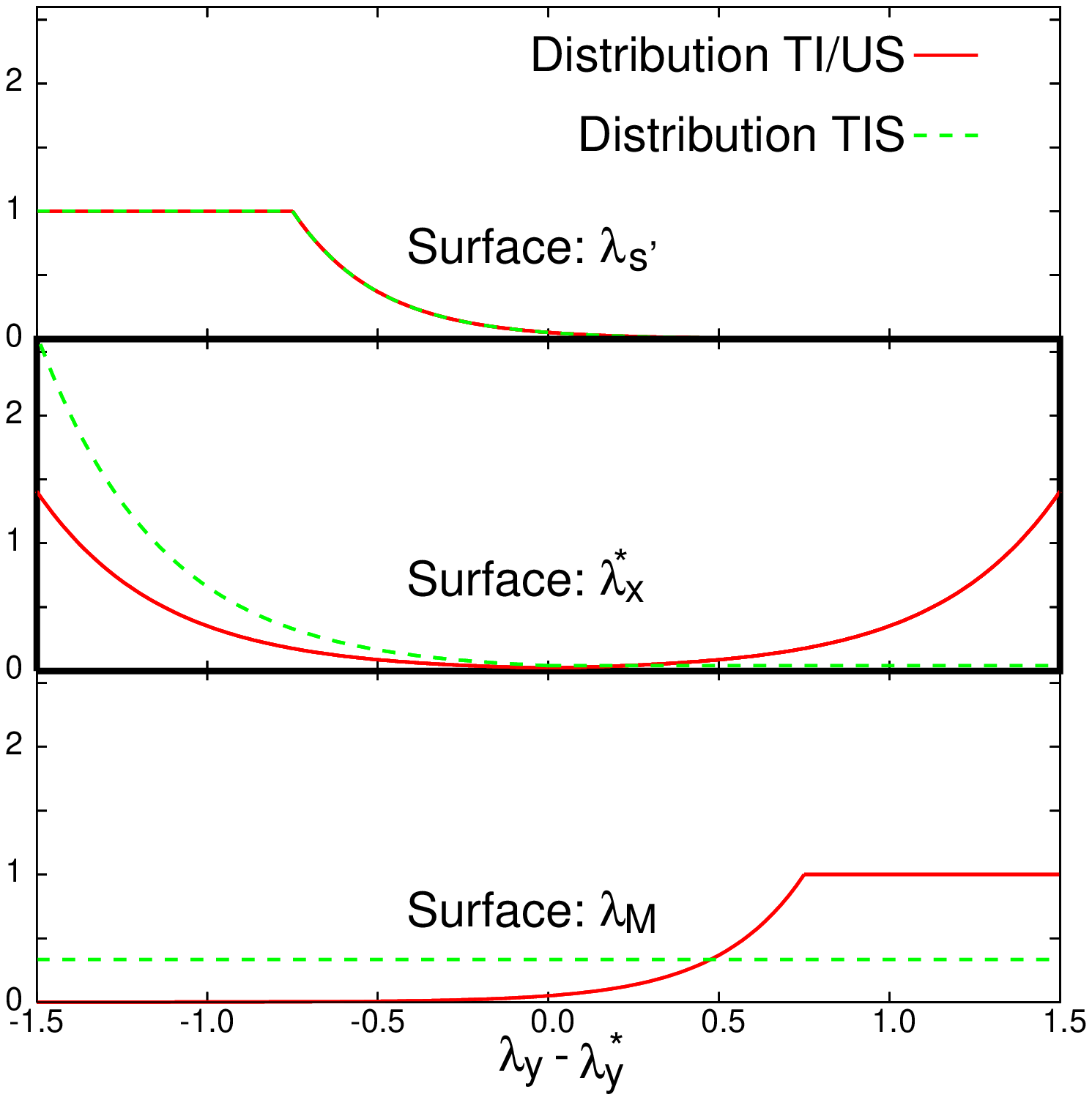}
\caption{(Color online) Distributions along the $\lambda_y$
coordinate for three surfaces and the TI/US and
TIS methods.
}
\label{distTITIS}
\end{center}
\end{figure}

Consider the potential defined by Eq.~(\ref{V2D})  
and the hypersurfaces $\lambda_{s'}, \lambda_x^*$ and $\lambda_M$ as depicted in Fig.~\ref{top2D}. Suppose that we apply  TI  or US using small windows located at these surfaces. 
The distribution at these surfaces along the $\lambda_y$ direction is then given by
\begin{align}
P_{\lambda_s}^{\rm TI/US}(\lambda_y)=
\frac{\lo\delta(\lambda_y) \delta(\lambda(x)-\lambda_s)  \rc}{\lo \delta(\lambda(x)-\lambda_s) \rc}
\label{distRF}
\end{align} 
where $\lambda_s$ can be either $\lambda_{s'}, \lambda_x^*$ and $\lambda_M$.
In TIS, we could look at the distribution of first crossing points with the $\lambda_s$ interface. 
This distribution is given by ($v=\dot{\lambda}_x(x_0)$)
\begin{align}
P_{\lambda_s}^{\rm TIS}(\lambda_y)=
\frac{\lo \delta(\lambda_y) v  \delta(\lambda(x)-\lambda_s) h_{0,s}^b  \rc}{\lo  v  \delta(\lambda(x)-\lambda_s) h_{0,s}^b  \rc}.
\label{distTIS}
\end{align} 
The additional $v$-term in the nominator and denominator of Eq.~(\ref{distTIS})
compared to Eq.~(\ref{distRF}) is due to the fact that TIS looks at crossing points
while the free energy distribution looks at points on the surface. A crossing point is a point that can cross the surface in a single timestep and because 
$\lim_{\Delta t \rightarrow 0} \frac{1}{\Delta t} 
\theta(\lambda_s-\lambda(x_0))  
\theta(\lambda(x_{\Delta t})-\lambda_s)=\delta(\lambda(x_0)-\lambda_s) 
\dot{\lambda}(x_0)$~\cite{titusthesis}
the additional $v$-term appears.  The other term $h_{0,s}^b$ in Eq.~(\ref{distTIS}) that is missing in
Eq.~(\ref{distRF})  ensures that not all velocities are considered. Starting from $x$ going backward 
in time, $\lambda_0$ should be hit before $\lambda_s$. This implies that we can write
$h_{0,i}^b=\theta(v)$ if $\lambda_\perp < 0$ and $h_{0,i}^b=\theta(v-\sqrt{2(H-V(\lambda_s,\lambda_y))/2})$ 
for $\lambda_\perp > 0$. Substituting this in Eq.~(\ref{distTIS}) yields
\begin{align}
P_{\lambda_s}^{\rm TI/US}(\lambda_y)=\frac{e^{-\beta V(\lambda_s,\lambda_y)}}{\int \ud
\lambda \, e^{-\beta V(\lambda_s,\lambda)}}, \,
P_{\lambda_s}^{\rm TIS}(\lambda_y)=\frac{e^{-\beta V'(\lambda_s,\lambda_y)}}{\int \ud
\lambda \, e^{-\beta V'(\lambda_s,\lambda)}} \label{TITIS}
\end{align}
with $V'(\lambda_x,\lambda_y)=V(\lambda_x,\lambda_y) \theta(-\lambda_\perp) 
+ H \theta(\lambda_\perp)$. 
Fig.~\ref{distTITIS} shows the distributions of Eqs.~(\ref{TITIS}) for three interfaces
for the case $\theta=33.7$, $R_y=1.5, W = 3.6$ and $\beta H=9$.
At the first interface in Fig.~\ref{distTITIS}, the two sampling distributions of Eqs.~(\ref{TITIS}) are exactly 
the same.  The distribution is maximized at the left side of $\lambda_y^*$. However, at the interface 
$\lambda^*_x$ the two distributions are very different. The TI/US distribution has now two maxima
at either side of $\lambda_y^*$.  As MD and most MC 
schemes generate a new phase points by making small displacements from the previous point,
the low probability region in the middle needs to be crossed to sample the distribution at either side.
However, as crossing this low probability region is a rare event on itself, the system  
might not cross this region during whole simulation  period. 
On the other hand, once the TI surface or US window is dragged over the barrier, the distribution is 
peaked at the right side of $\lambda_y^*$ as can been seen from the shape of the distribution at $\lambda_M$. 
If we move the surface back
from $\lambda_M$ to $\lambda_x^*$, the system would be stuck again, but now at the right side
of $\lambda_y^*$, which illustrates the hysteresis problem.  Only when the sampling is done extensively long, 
both maxima at the surface $\lambda_x^*$ can be sampled in a single simulation. However, all measurements between two crossing events 
are correlated and basically contribute as a single uncorrelated measurement. 
Hence, ${\mathcal N_C^{\rm TI/US}}$ becomes  exceedingly large.

TIS does not have this problem. The distribution at $\lambda_x^*$ has still only one maxima. 
The distribution becomes uniform at $\lambda_M$.  The divergence of 
${\mathcal N_C^{\rm TIS}}$ is, hence, not to be expected. Only when there are distinctive different
reaction channels, ergodic sampling becomes difficult as for any method. 
Still the sampling of multiple reaction channels is likely more effective using path sampling 
than TI or US due to the non-locality
of the shooting move~\cite{Geissler99}.  These findings and the results of~\ref{TIS2D}
actually prove the relative insensitivity of TIS to the RC 
as compared to the RF methods.
This quality  has been assigned to path sampling methods before, but
to our knowledge, this is the first time that this is rigorously proven for reaction rate calculation
methods. 
This points out a significant advantage of TIS for systems for which a proper RC can not easily be derived.

It is important to note that other path sampling approaches such as PPTIS (or Milestoning)
and FFS do not have this advantage. The PPTIS approximation fails 
for $\theta>0$ except when the interfaces are very far apart. For instance,
consider the interfaces $\lambda_{s-1}<\lambda_{s}<\lambda_{s+1}$
with $\lambda_{s}=\lambda_x^*$. The PPTIS memory-loss assumption
reveals that the trajectories that start at $\lambda_{s-1}$ and cross
$\lambda_{s}$ have on average the same probability to reach 
$\lambda_{s+1}$ as the trajectories that start at $\lambda_{0}$ and cross
$\lambda_{s}$.
This can only be valid if $\lambda_{s-1} \leq  \lambda_{s'}$ (See Fig.~\ref{top2D}).
If $\lambda_{s-1}$ is set closer to $\lambda_s$, many trajectories that cross
$\lambda_{s-1}$ along a lines $\lambda_y < \lambda_y^*$ 
will not come from $\lambda_0$ if they are followed 
backward in time and the PPTIS
approximation results in an overestimation of the reaction rate. 
In turn, the large interface distances results that some of the PPTIS 
crossing probabilities are very small 
and can only be determined efficiently using TIS. 

FFS, although in principle exact, also has a problem  for non-zero angles $\theta$. 
As FFS only propagates forward in time
using the successful paths from the previous simulation,  
all the simulations  become correlated. This implies that whenever the simulation at the first interface
is in error,
all other simulation results will be erroneous as well even if these simulations are performed infinitely long. This is a direct effect of the non-validity
of Eq.~(\ref{assum_unc}) for FFS. 
Coming back to Fig.~\ref{distTITIS},
the flat distribution at $\lambda_M$
can only be reproduced using FFS when at the previous 
interfaces (as $\lambda_x^*$ and $\lambda_{s'}$) 
a significant number of pathways is sampled in low-populated 
right tail of the path distribution which requires the 
sampling of an  extremely large number of
pathways.

\section{conclusions}
\label{seccon}
We have derived analytical expressions to determine the efficiency of 
different computational methods for the calculation of reaction rates.
The efficiency has been expressed as the computational cost to obtain an 
overall relative error of 1 when all algorithmic parameters are optimized. 
We have called this property the CPU efficiency time $\tau_{\rm eff}$. 
In App.~\ref{secect}, we have derived the  
CPU efficiency for very general cases. 
This also reveals an important generic result of how one should
divide a fixed total simulation time  
among a set of
independent simulations
to get the lowest overall error. 
After a first round of simulations,  reasonable estimates
of $\tau_{\rm eff}^{(s)}$ can be obtained.
Then,
the minimal additional simulation time, needed
in each simulation, to obtain the overall best performance
can be calculated  and 
used for a second round of simulations. This approach can be very profitable
and it is not restricted to rate calculations.
 
We have applied these rules to determine the efficiency of a
simple 2D benchmark system that allows an analytical approach. This
offers a way to 
compare the efficiency of the different methods in a very transparent
way. The two classes of methods that we compared are the RF 
methods and the path sampling methods. The  RF methods
consists of the calculation of the free energy barrier and the
calculation of the transmission coefficient. Both for the free energy
as for the transmission coefficient calculation different methods
exist.  
For the free energy calculation we compared two approaches
of US. One using a series of rectangular windows and one using a
single optimal biasing potential.  
For the path sampling methods, we have concentrated on the TIS technique 
which is an improvement upon the original TPS algorithm to calculate rates.
The PPTIS approach reduces the computational cost even more
but relies on the 
approximation of 'memory loss'. 
As for US, we suggest that a single bias potential based on $\lambda_{\max}[X]$
could replace the discrete set of interfaces. 

The $\theta=0$ case corresponds to the situation where the chosen RC
is optimal. The TST approximation is then
exact
so that the free energy calculation is sufficient for the RF methods.
We found an efficiency 
scaling equal to $(\beta H)^2$ for US using rectangular windows.  We
obtained the same scaling rules for standard TIS.  
Using a single continuous bias potential, the US and TIS 
efficiency can, in the optimal
case, become independent of the barrier height and temperature.  
However, knowing the optimal bias basically implies 
that one already knows the answer. US using several windows and standard TIS
are more robust approaches if no a-priori knowledge of
the system is available.
This shows that TIS and US
compare very well in efficiency for all barrier heights and
temperatures and that the difference can only rely in a
single prefactor.
It is likely that US is a bit better than TIS
that has to be faithful to the natural dynamics of the system.
US has more flexibility to optimize the method such as choosing the
best MC moves that minimizes the number of correlations.
 
When $\theta>0$, the chosen RC, $\lambda_x$, is no longer optimal.
In contrast to the $\theta=0$ case, the principal computational effort
of the RF method lies in the calculation of the dynamical correction. We showed that
the BC method scales as  $\sim \exp(4 \beta H \sin \theta R_y/W)$, while the BC2 and EPF methods are quadratically 
more efficient. 
The EPF method is, however, a bit more than a factor 2 more 
faster than the BC2 method. The exponential dependence 
on $ \beta H \sin \theta$ indicates
that a small deviation from the optimal RC can have dramatic consequences 
for the efficiency of the
RF methods if they are applied to high barrier or low temperature systems.
In contrast, the TIS efficiency scaling is only $\sim (\beta H \sin \theta R_y/W)^2$. 
We also discuss the potential problem of hysteresis in US and TI 
when a non-optimal RC is chosen and why 
this problem does not occur for TIS for the 2D barrier. This gives another evidence that the TIS method is less sensitive the choice of RC. 

The advantage that path sampling does not require a RC has been advocated many times. However,
although
this statement is quite evident if the main object is to sample a representative 
set of reactive trajectories, it is not so evident for the calculation of 
reaction rates.
The calculation of the reaction rate 
still requires a RC (the only exception we know of is the method proposed 
in Ref.~[\onlinecite{ErpBol2004}] 
using the pathlength as transition parameter). 
However, we are now the first to prove  that 
a path sampling-based reaction rate calculation method, using TIS,  
is potentially orders of magnitude 
faster than the RF-based methods whenever the right RC can not be determined. 
The reason is because TIS  uses an importance sampling technique to calculate the dynamical factor.
The importance sampling of the
RF methods only concentrates on the 
free energy. Therefore, whenever the dynamical factor is low (e.g. due to a wrong choice of RC), 
these methods automatically run into problems. 
The main question remains
whether it is more profitable to search for a good RC and use the RF methods, or take a simple
order parameter (non-optimal RC) and use the TIS method. This question has not an easy
answer and depends on the complexity of the system.

I would like to thank Daniele Moroni for fruitful discussions and 
carefully reading this paper. I am also grateful to Rosalind Allen
who made useful suggestions to improve the first version of this paper.
This work was support by a Marie Curie Intra-European Fellowships
(MEIF-CT-2003-501976) within
the 6th European Community Framework Programme.


\appendix

\section{General definitions\label{sgn}}

\subsection{Ensembles averages in phase and path space}
\label{subsensem}
We denote with $x$ a phase space point $\{r,p\}$ where $r$ are the
Cartesian coordinates and $p$ the momenta.  The examples presented
here consider only 1 particle in a two-dimensional potential,
but $x,r,p$ are in general multidimensional vectors.  The distribution
of $x$ is given by the probability density $\rho(x)$. In case of
Boltzmann statistics $\rho(x) \propto e^{-\beta E(x)}$ with $E(x)$ the
total energy at phase point $x$ and $\beta=1/(k_B T)$ with $T$ the
temperature and $k_B$ the Boltzmann constant.  Suppose $a$ is the
value of a quantity we want to compute. In many cases such a quantity
equals the expectation value of a certain observable.  We denote
this observable with $\hat{a}(x)$ which is a function of $x$.  Then
the expectation value or population mean $\lo \hat{a} \rc$ is given by
\begin{align}
 \lo \hat{a}(x) \rc=\frac{\int \mathrm{d}x \, \hat{a}(x) \rho(x)}{\int
 \mathrm{d}x \, \rho(x)}
\label{phaseav}
\end{align}
and $a=\lo \hat{a} \rc$ for this specific case.  Path sampling
simulations require a more extensive formulation especially when
stochastic dynamics is applied. The discrete representation of a path
is the most convenient, where a path $X$ is defined as a set of
$\tau^b+\tau^f+1$ successive phase points that determine the system at
intervals of $\Delta t$ along a certain trajectory: $X=\{ x_{-\tau^b
\Delta t},x_{(1-\tau^b) \Delta t},\ldots, x_{-\Delta t};x_{0};
x_{\Delta t},\ldots , x_{+\tau^f \Delta t}, \}$.  In TIS, the backward
$\tau^b$ and forward $\tau^f$ time indices are not fixed, but depend
on when a certain interface is crossed. The weight of the path is
given by the initial distribution at time $t=0$ and the probability
densities corresponding to the history and future of the path:
\begin{align}
P[X]=\rho(x_0) &\prod_{i=1}^{\tau^b} p_{\rm n}(x_{(1-i)\Delta t}
\leftarrow x_{-i \Delta t})\nonumber \\ \times &\prod_{i=1}^{\tau^f}
p_{\rm n}(x_{(i-1)\Delta t}\rightarrow x_{i \Delta t} )
\label{pathweight1} 
\end{align}
where $p_{\rm n}(x \rightarrow y)$ is the probability that the natural
dynamics of the system brings you from $x$ to $y$ given you are in $x$, and
$p_{\rm n}(x \leftarrow y)$ is the probability
that if you are in $x$ you came from $y$ in the past.
As shown in Ref.~[\onlinecite{ErpMoBol2003}], 
whenever the system is in a steady state,
Eq.~(\ref{pathweight1}) is identical to
 \begin{align}
P[X]=\rho(x_{-\tau^b \Delta t})\prod_{i=1}^{\tau^b+\tau^f}& p_{\rm n}(x_{(i-\tau^b-1)\Delta t}
\rightarrow  x_{(i-\tau^b)\Delta t} )
\label{pathweight2} 
\end{align}
which corresponds to the path weight as expressed in the original TPS papers~\cite{Dellago02} 
with the only difference
that the starting index is $-\tau^b$ instead of 0. 
Now, if $\hat{a}[X]$ is a function defined in path space, the population mean is given by
\begin{align}
\lo \hat{a}[X] \rc=\frac{\int { \mathcal D}X \, \hat{a}[X] P[X]}{ \int {\mathcal D}X \, P[X]},
\label{pathav}
\end{align}
with ${\mathcal D}X=\prod_{i=-\tau^b}^{\tau^f} dx_{i \Delta t}$ and $P[X]$ given by~Eq.(\ref{pathweight1}). 

In the following, we use $q$ as 
a point that is defined in either phase or path space, i.e. $q$ can either represent $x$ or $X$.  
Besides the simple ensemble averages 
(\ref{phaseav}) and (\ref{pathav}), we can also define the biased ensemble 
average 
$\lo \hat{a} \rc_\Omega$ using a weight 
function $\Omega(q)$. This biased ensemble is defined as
\begin{align}
\lo \hat{a}(q) \rc_\Omega \equiv \frac{\lo \hat{a}(q) \Omega(q) \rc}{\lo \Omega(q) \rc}.
\label{wens}
\end{align} 
In practice, sampling the biased ensemble $\lo \ldots \rc_\Omega$ by means of MD 
simply requires the addition of a term $-(\ln \Omega)/\beta$
to the potential of the system.  
In MC, this is achieved by a change of the acceptance-rejection criterion 
from $\min[1,\rho(x^{(n)})/\rho(x^{(o)})]$  to  $\min[1,\rho(x^{(n)})\Omega(x^{(n)})/(\rho(x^{(o)})\Omega(x^{(o)}))]$ 
with $x^{(n)}$ and  $x^{(o)}$ the new and old generated MC
points~\cite{FrenkelSmit}.

Ensemble averages in path space $\lo a[X] \rc$ can depend on the type
of dynamics (hence, on the hopping probabilities $p_{\rm n}$). 
To specify this dependency, explicitly, we use $\lo \ldots \rc_{;p_{\rm n}}$,
which allows to generalize these path ensemble averages to arbitrary dynamics.  
The hopping probabilities of  a simulation method $p_{\rm s}(x \rightarrow y)$ can  be of any
type, for instance Monte Carlo, 
and do not need to be related to the 
natural dynamics of the system. 
Therefore, these ensemble averages are annotated by
$\lo \ldots \rc_{;p_{\rm s}}$. When $;p$ is not specified, we assume
that the natural dynamics is applied or that the property is independent of
the hopping probability $p$. Hence,  in our notation
$\lo a[X] \rc=\lo a[X] \rc_1=\lo a[X] \rc_{;p_{\rm n}}=
\lo a[X] \rc_{1;p_{\rm n}}$.

\subsection{Standard deviation, variance, covariance,  and correlation}
\label{subscor}
The standard deviation of $\hat{a}$ is defined as
\begin{align}
\sigma_{\hat{a}} &= \sqrt{\lo (\hat{a} -\lo \hat{a} \rc )^2 \rc}=\sqrt{\lo \hat{a}^2 \rc -\lo \hat{a} \rc^2} \nonumber \\
&= \sqrt{\lo (\hat{a} -a )^2 \rc} = \sqrt{\lo \hat{a}^2 \rc -a^2}.
\label{std}
\end{align} 
Related to the standard deviation is the
variance of $\hat{a}$
\begin{align}
{\rm Var}(\hat{a}) \equiv \lo (\hat{a} -\lo \hat{a} \rc )^2 \rc=
\sigma_{\hat{a}}^2
\end{align}
and the covariance of two functions $\hat{a}$ and $\hat{b}$
\begin{align}
{\rm Cov}(\hat{a},\hat{b})& \equiv  \lo (\hat{a} -\lo \hat{a} \rc )(\hat{b} -\lo \hat{b} \rc ) \rc \nonumber \\
&=  \lo \hat{a} \hat{b} \rc -ab  
\label{defcov}
\end{align}
with ${\rm Cov}(\hat{a},\hat{a})={\rm Var}(\hat{a})$.
A simulation $s$, which can either be MC, MD, or TPS/TIS generates a set of points 
$\{ q_0, q_1, q_2, \ldots , q_n \}$ in either phase or path space. 
Each point $q_i$ is simulated with a certain probability $\rho_s(q_i)$ and the chance that
after $q_i$ another point $q_{i+1}$ is sampled is given by $p_{\rm s}(q_i \rightarrow q_{i+1})$. 
We denote $a_i=\hat{a}(q_i)$. Now, the sample mean $\bar{a}(n)$ 
for a simulation of length $n$ is given by
\begin{align}
\bar{a}(n) \equiv \frac{1}{n} \sum_{i=1}^n a_i.
\label{sammean}
\end{align} 
Eq.~(\ref{sammean}) converges to the exact value, $\bar{a}(n)=a $, 
in the limit $n\rightarrow \infty$.
The standard deviation in the mean $\sigma_{\bar{a}(n)}$ is defined 
as the standard deviation in the
set of points  $\{ \bar{a}^{(1)}(n),  \bar{a}^{(2)}(n), \bar{a}^{(3)}(n), \bar{a}^{(4)}(n), \ldots\}$ 
that is obtained after performing a large number of independent simulations, $s=1, 2, 3, \ldots$, 
each of length $n$ and with result  $\bar{a}^{(s)}(n)$.
Hence, we can write 
\begin{align}
\sigma_{\bar{a}(n)}^2=\lo ( \bar{a}(n)-a  )^2 \rc_{;p_{\rm s} }
=\frac{1}{n^2} \sum_{i=1}^{n} \sum_{j=1}^n \lo (a_i- a  )(a_{j}- a   )
\rc_{;p_{\rm s}}
\nonumber \\
=  \frac{1}{n^2} \Big\{ \sum_{i=1}^{n}  \lo (a_i- a  )^2  \rc_{;p_{\rm s}}
+2 \sum_{i=1}^{n-1} \sum_{j=i+1}^{n} \lo (a_i- a  )(a_{j}- a   )
\rc_{;p_{\rm s}} \Big\} \label{A1}.
\end{align}
Using time translation invariance, we can show that 
$\lo (a_i- a  )^2  \rc_{;p_{\rm s}}=\lo (a_0- a  )^2
\rc_{;p_{\rm s}}
= \lo (\hat{a}- a  )^2  \rc$ and  $\lo (a_i- a  )(a_{j}- a   )
\rc_{;p_{\rm s}}
=\lo (a_0- a  )(a_{j-i}- a   ) \rc_{;p_{\rm s}}$.
Now, we assume an exponentially decay in
correlation for large $l\equiv j-i$ which yields for
large $n$:
\begin{align}
\sigma_{\bar{a}(n)}^2 & \approx   \frac{1}{n} \Big\{
\lo (\hat{a}- a  )^2  \rc
+2 \sum_{l=1}^{\infty} \lo (a_0- a  )(a_{l}- a   )
\rc_{;p_{\rm s}} \Big\}
\nonumber \\
&= \frac{1}{n} \Big\{   \sigma_{\hat{a}}^2
\Big[ 1 + 2 \sum_{l=1}^\infty
\frac{  \lo (a_0- a  )(a_{l}- a   )
\rc_{;p_{\rm s}}  }{ \lo (\hat{a}- a  )^2 \rc }  \Big] \Big\}
\label{stdm2}
\end{align}
which gives     
\begin{align}
\sigma_{\bar{a}(n)}
&=   \frac{\sigma_{\hat{a}}}{\sqrt{n}} \sqrt{ 1 + 2 n_c^{aa}}  
\label{stdm}.
\end{align}
In Eq.~(\ref{stdm}), $n_c^{aa}$ is the correlation number or the integrated auto-correlation function,
which is a special case of $n_c^{ab}$, defined as
\begin{align}
n_c^{ab} &\equiv   \sum_{l=1}^\infty C_{ab}(l) \label{ncab}
\end{align}
with
\begin{align}
C_{ab}(l) &\equiv
\frac{ \lo (a_0- a )(b_{l}- b   ) \rc_{;p_{\rm s}} }{\lo (\hat{a}- a  )
(\hat{b}- b  ) \rc  }.
\label{corfun}
\end{align} 
If two functions $\hat{a}$ and $\hat{b}$ are uncorrelated, 
$\lo \hat{a}\hat{b} \rc =\lo \hat{a} \rc \lo \hat{b} \rc=ab$. 
Hence, if the simulation data are not correlated  $p_{\rm s}(q_i \rightarrow q_{i+1})
=\rho_s(q_{i+1})$  and 
$ \lo (a_i- a  )(a_{i+l}- a ) \rc
= \lo (a_i- a  )\rc \lo (a_{i+l}- a ) \rc=0$ and  $n_c^{aa}=0$.

\subsection{Statistical errors and propagation rules}
\label{subserr}
Let  $\Delta a$ be the absolute error of a quantity $a$
and let $\delta a \equiv \Delta a/a$ be the
relative error. For a sequence of measurements 
$\{a_0,a_1,\ldots, a_n\}$, the absolute error 
in the average $\bar{a}(n)$ is defined as
the standard deviation in the mean:
\begin{align}
\Delta a(n) \equiv \sigma_{\bar{a}(n)}.
\label{deferr}
\end{align}
Using Eq.~(\ref{stdm}),
the absolute and relative errors are 
\begin{align}
\Delta a(n) = \sigma_{\hat{a}} \sqrt{\frac{1+2 n_c^{aa}}{n}}, \quad
\delta a(n) = \frac{\sigma_{\hat{a}}}{a} \sqrt{\frac{1+2 n_c^{aa}}{n}}.
\label{errs}
\end{align}
This shows that it is of eminent importance for the efficiency of the method 
that the trajectories generated 
by the computer algorithm minimize 
the correlation number $n_c$ as much as possible. 
In MC, this gives rise to conflicting strategies.
In general MC algorithms
select a new 
phase point by making a random displacement 
from a previous point, which is then 
accepted or rejected. When the displacement is small, the new point resembles
strongly to the old one which can be a 
source of correlations. On the other hand, if the randomized 
displacement is too large, it is likely rejected after which the old phase 
point is counted again. Therefore, a high rejection probability also
introduces correlations.  
The optimum maximum displacement is a compromise between these two
effects. 
The creation of a single step in path sampling is much more expensive than
standard MC or MD, but, on the other hand, 
the correlation number $n_c$ is usually 
much lower.

Suppose a quantity $a$ is not equal to the  expectation value of
a single operator, but,  for instance,  depends on two other quantities $b$
and $c$ via a function $f$: $a=f(b,c)$. The error in $a$ can then be determined 
using the error propagation rules. 
Say $\bar{b}(m)$ and  $\bar{c}(n)$ are the approximations of $b$ and $c$ 
after $m$ and $n$ simulation cycles respectively. The resulting error 
$\Delta a(m,n)$ can then be derived as follows.   
As
$(\bar{b}(m)-b)$ and $(\bar{c}(n)-c)$ are small for large
values of $m$ and $n$, we can invoke following Taylor expansion:
\begin{align}
\bar{a}(m,n) &\equiv f( \bar{b}(m) , \bar{c}(n)) \nonumber \\
 &\approx  f( b ,  c )+ \frac{\partial f}{\partial b} ( \bar{b}(m)-b)+
\frac{\partial f}{\partial c} ( \bar{c}(n)-c). \label{taylor}
\end{align}
As in Eq.~(\ref{deferr}) the error $\Delta a(m,n)$ equals the standard deviation
of the mean $\bar{a}(m,n)$.
Hence,
\begin{align}
\Delta a^2(m,n) \equiv \lo (\bar{a}(m,n)-a)^2
\rc_{;p_{\rm s}}
= \Big(\frac{\partial f}{\partial b}\Big)^2
\lo (\bar{b}(m)-b)^2 \rc_{;p_{\rm s}} \nonumber \\
+
  \Big(\frac{\partial f}{\partial c}\Big)^2
\lo (\bar{c}(n)-c)^2 \rc_{;p_{\rm s}}
+ 2 \frac{\partial f}{\partial b}  \frac{\partial f}{\partial c}
\lo  (\bar{b}(m)-b )(\bar{c}(n)-c) \rc_{;p_{\rm s}}.
\label{damn2}
\end{align}
Substitution of Eq.~(\ref{defcov}), (\ref{deferr}) and (\ref{stdm})
yields 
\begin{align}
\Delta a^2(m,n) 
&=
\Big(\frac{\partial f}{\partial b}\Big)^2 \Delta b^2(m)+ 
  \Big(\frac{\partial f}{\partial c}\Big)^2 \Delta c^2(n)    \nonumber \\
&+ 2 \frac{\partial f}{\partial b}  \frac{\partial f}{\partial c}
{\rm Cov}\big(\bar{b}(m),\bar{c}(n)\big).
\label{DAmn}
\end{align}
Then, from
$\Delta a=f(b,c) \delta a$,
$\Delta b=b \delta b$, $\Delta c=c \delta c$ we get the propagation rule for the relative error
\begin{align}
\delta a^2(m,n) &=  \big[b \frac{\partial f}{\partial b}/f \big]^2 \delta b^2(m)+
  \big[c \frac{\partial f}{\partial c}/f \big]^2 \delta c^2(n)
 \nonumber \\
  &+2 \big[ \frac{\partial f}{\partial b}\frac{\partial f}{\partial c}/f^2
\big] {\rm Cov}(\bar{b}(m),\bar{c}(n)).
\label{da2}
\end{align}
Eqs.~(\ref{DAmn}) and (\ref{da2}) are the most general 
rules for the propagation 
of errors when $a$ depends on two quantities $b$ and $c$. It can 
straightforwardly be 
generalized to more arguments such that $a=f(b,c,d,\ldots)$.

\section{CPU efficiency times }
\label{secect}

\subsection{CPU efficiency time for a single ensemble average}
\label{subssa}
Let us define $\tau_{\rm cyc}$ as the average CPU time to perform 
a single MC, MD or path sampling cycle.   
Moreover, we denote $\tau_{\rm sim}=n \tau_{\rm cyc}$ as the total  
CPU simulation time after $n$ cycles.
We introduce now the CPU efficiency time $\tau_{\rm eff}(a)$ 
to obtain a relative error  $\delta a$
equal to one. Putting $\delta a=1$ in Eq.~(\ref{errs}) and using 
$n = \tau_{\rm sim} /\tau_{\rm cyc}=\tau_{\rm eff} /\tau_{\rm cyc}$ 
gives
\begin{align}
\tau_{\rm eff}(a)=\Big( \frac{\sigma_{\hat{a}}}{a} \Big)^2 (1+2 n_c^{aa}) 
\tau_{\rm cyc}.
\label{taueff}
\end{align}
This is the general CPU efficiency time to obtain a relative error of 1
in a single simulation average.  This CPU efficiency time is a characteristic 
property of the 
quantity $a$  (actually of the whole function $\hat{a}$) 
and the simulation method 
(via the $p_{\rm s}$ hopping probabilities and $\tau_{\rm cyc}$). 
Whenever $\tau_{\rm eff}(a)$ is known
the absolute and relative errors after a simulation period $\tau_{\rm sim}$
are directly given by: 
\begin{align}
\Delta a = a \sqrt{\tau_{\rm eff}(a)/\tau_{\rm sim}}, \quad 
\delta a = \sqrt{\tau_{\rm eff}(a)/\tau_{\rm sim}}.
\label{erreff}
\end{align}
If $\hat{a}$ is a binary operator such that $\hat{a}(q)$ is either 1 or 0,
then $\hat{a}^2=\hat{a}$ in Eq.~(\ref{std}) and Eq.~(\ref{taueff}) equals 
\begin{align}
\tau_{\rm eff}(a)=\frac{1-a}{a}(1+2 n_c^{aa})\tau_{\rm cyc}.
\label{taubin}
\end{align}
Therefore, the $\tau_{\rm eff}(a)$ becomes very large for small values of $a$ 
and, hence, an accurate evaluation becomes problematic. 

\subsection{CPU efficiency time for a composite quantity}
\label{subscq}
The CPU efficiency time for a composite value  $a=f(b,c)$, where $b=\lo \hat{b} \rc$ and 
$c=\lo \hat{c} \rc$ are determined simultaneously in a 
single simulation, can be 
derived 
following the
the same lines as in Eq.~(\ref{stdm2}) for the covariance 
\begin{align}
&{\rm Cov}(\bar{b}(n),\bar{c}(n)) = \frac{1}{n^2} \Big\{ \sum_{i=1}^n \lo (b_i-b)(c_i-c) \rc_{;p_{\rm s}} \nonumber \\
&+ \sum_{i=1}^n \sum_{j=i+1}^n [ \lo (b_i-b)(c_j-c)
\rc_{;p_{\rm s}}
+ \lo (c_i-c)(b_j-b)
\rc_{;p_{\rm s}} ] \Big\} \nonumber \\
 &\approx \frac{1}{n} \Big\{ \lo (\hat{b}-b)(\hat{c}-c) \rc \times
  \Big[  1 +
  \nonumber \\
   & \sum_{l=1}^\infty \frac{ \lo (b_0-b)(c_l-c)
\rc_{;p_{\rm s}}
+ \lo (c_0-c)(b_l-b) \rc_{;p_{\rm s}} }{
\lo (\hat{b}-b)(\hat{c}-c) \rc
} \Big] \Big\} \nonumber \\
& = \frac{1}{n} {\rm Cov}(\hat{b},\hat{c}) \Big[ 1 + n_c^{bc}+n_c^{cb} \Big]
\label{Cov2}.
\end{align}
Here, we inserted Eq.~(\ref{defcov}) and (\ref{corfun}) in the last line.
Substitution of Eqs.(\ref{erreff}) and (\ref{Cov2}) into Eq.~(\ref{da2})
and using $n=\tau_{\rm sim}/\tau_{\rm cyc}$ gives
\begin{align}
\delta a^2 &=  \big[b \frac{\partial f}{\partial b}/f \big]^2 \frac{\tau_{\rm eff}(b)}{\tau_{\rm sim}}+
  \big[c \frac{\partial f}{\partial c}/f \big]^2 \frac{\tau_{\rm eff}(c)}{\tau_{\rm sim}}
\label{da3} \nonumber \\
  &+2 \big[ \frac{\partial f}{\partial b}\frac{\partial f}{\partial c}/f^2
\big] {\rm Cov}(\hat{b},\hat{c}) \big\{1 + n_c^{bc}+n_c^{cb} \big\} \frac{\tau_{\rm cyc}}{\tau_{\rm sim}}.
\end{align}
Taking $\delta a^2=1$ and $\tau_{\rm sim}=\tau_{\rm eff}(a)$
directly results in
\begin{align}
 \tau_{\rm eff}(a) &\equiv  \big[b \frac{\partial f}{\partial b}/f \big]^2 \tau_{\rm eff}(b)+
  \big[c \frac{\partial f}{\partial c}/f \big]^2 \tau_{\rm eff}(c) 
\label{CPUcomp} \\
  &+2 \big[ \frac{\partial f}{\partial b}\frac{\partial f}{\partial c}/f^2 \big] {\rm Cov}(\hat{b},\hat{c})
  \Big\{ 1+ n_c^{bc}+n_c^{cb} \Big\} \tau_{\rm cyc} \nonumber
 \end{align}
 where $\tau_{\rm eff}(b)$ and $\tau_{\rm eff}(c)$ are given by Eq.~(\ref{taueff}).
For example,  if $a=b^i c^j$, the efficiency time of $a$ equals
\begin{align}
 \tau_{\rm eff}(a) &= i^2 \tau_{\rm eff}(b)+j^2 \tau_{\rm eff}(c) \nonumber \\
 &+ 2 \frac{ij}{bc} {\rm Cov}(\hat{b},\hat{c})
  \Big\{ 1+ n_c^{bc}+n_c^{cb} \Big\} \tau_{\rm cyc}.  \label{taucomij}
\end{align}

\subsection{CPU efficiency time for a series of simulations}
\label{subsser}
Suppose $a=f(b,c)$ and $\bar{b}(m)$ and $\bar{c}(n)$ are obtained
via two different simulations. 
Hence, $n$ and $m$ are not necessarily the same. In the following we assume that the different 
simulations are uncorrelated.
Thus, we assume that for two different simulations (1) and (2) the 
following holds
\begin{align}
{\rm Cov}( \bar{b}(m),\bar{c}(n))=
\lo (\bar{b}(n) -b  )(\bar{c}(m) -c ) \rc_{;p_{{\rm s}(1,2)}}= 0.
\label{assum_unc}
\end{align}
Here, the subscript $;p_{{\rm s}(1,2)}$ indicates that the ensemble average can depend
on how the two simulations are connected.
Assumption~(\ref{assum_unc}) is, for instance, true for US when two independent simulations are performed using two overlapping windows. It also holds for  TIS where the outcome of an interface sampling simulation at a certain interface is independent of the result of the previous interface simulation results.   Eq.~(\ref{assum_unc}) does not hold for 
the most common implementation of the reactive flux method.
In this approach, the importance sampling to determine the free energy barrier
is simultaneously used to obtain a representative set of configuration points at the TS dividing
surface.
These configurations with randomized Gaussian distributed velocities
initiate the dynamical trajectories that 
determine the transmission coefficient~\cite{FrenkelSmit,Strnad}.
Moreover, at variance with TIS,  Eq.~(\ref{assum_unc}) is not true for the Forward Flux Sampling
(FFS) method, that was devised by Allen \emph{et al.}~\cite{FFS,FFS2}. Here, the result of the 
interface sampling at one interface depends on the results of all the previous interfaces.
The importance of Eq.~(\ref{assum_unc}) is further discussed in Sec.~\ref{susechys}.
 
If the total simulation time 
$\tau_{\rm sim}(a) = \tau_{\rm sim}(b) + \tau_{\rm sim}(c)=m \tau_{\rm cyc}(b) + n \tau_{\rm cyc}(c)$ 
is fixed, we still have some freedom in choosing $n$ and $m$ or, equivalently,
choosing $\tau_{\rm sim}(b)$ and $\tau_{\rm sim}(c)$.
First, by substitution of Eqs.~(\ref{erreff}) and (\ref{assum_unc})
into Eq.~(\ref{da2}) we obtain
\begin{align}
\delta a^2=
 \big[ b \frac{\partial f}{\partial b}/f \big]^2
 \frac{\tau_{\rm eff}(b)}{\tau_{\rm sim}(b)}
+ \big[ c \frac{\partial f}{\partial c}/f \big]^2
 \frac{\tau_{\rm eff}(c)}{\tau_{\rm sim}(c)}.
\label{d2a}
\end{align}
A logical approach, although not the optimum, is to
give each simulation the same simulation time. 
We will denote the efficiency time that results from this strategy
$\tau_{\rm eff}'(a)$. Taking
$\tau_{\rm sim}(b)
=\tau_{\rm sim}(c)=\frac{1}{2} \tau_{\rm sim}(a)$ for
$\tau_{\rm eff}'(a)=\tau_{\rm sim}(a)$ and  $\delta a^2=1$ 
gives
\begin{align}
\tau_{\rm eff}'(a)=
 2 \Big[ \big[ b \frac{\partial f}{\partial b}/f \big]^2
 \tau_{\rm eff}(b)
+ \big[ c \frac{\partial f}{\partial c}/f \big]^2
 \tau_{\rm eff}(c) \Big].
\label{teffa1}
\end{align}
Alternatively,  we could try to obtain the same error in each simulation.
The corresponding efficiency time will be annotated as $\tau_{\rm eff}''$.
Then, we need to use simulation times proportional to 
$\propto \tau_{\rm eff}(b), \tau_{\rm eff}(c)$.
Hence, we take
$\tau_{\rm sim}(b)=
\tau_{\rm sim}(a) \tau_{\rm eff}(b) /[ \tau_{\rm eff}(b)+ \tau_{\rm eff}(c) ]$
and
$\tau_{\rm sim}(c)=
\tau_{\rm sim}(a) \tau_{\rm eff}(c) /[ \tau_{\rm eff}(b)+ \tau_{\rm eff}(c) ]$ 
in Eq.~(\ref{d2a}) which results in
\begin{align}
\tau_{\rm eff}''(a)=
 \Big( \tau_{\rm eff}(b)+  \tau_{\rm eff}(c) \Big)   
\Big[ \big[ b \frac{\partial f}{\partial b}/f \big]^2
+ \big[ c \frac{\partial f}{\partial c}/f \big]^2 \Big].
\label{teffa2}
\end{align}
In order to determine the lowest $\tau_{\rm eff}$,
we add
Lagrange-multipliers constrains to Eq.~(\ref{d2a}) in order to fix the total
simulation time $\tau_{\rm sim}(a)$
\begin{align}
&\delta a^2[\tau_{sim}(b),\tau_{sim}(c),\eta] \equiv
 \big[ b \frac{\partial f}{\partial b}/f \big]^2
 \frac{\tau_{\rm eff}(b)}{\tau_{\rm sim}(b)} \nonumber \\
&+
 \big[ c \frac{\partial f}{\partial c}/f \big]^2
 \frac{\tau_{\rm eff}(c)}{\tau_{\rm sim}(c)}
-\eta^2 \big\{ \tau_{\rm sim}(a) - \tau_{\rm sim}(b)-\tau_{\rm sim}(c) \big\}
\label{Lagrange}
\end{align}
and minimize Eq.~(\ref{Lagrange}) with respect to all its arguments.
Taking the derivative to $\tau_{\rm sim}(b)$ gives
\begin{align}
 - \big[ b \frac{\partial f}{\partial b}/f \big]^2 \frac{\tau_{\rm eff}(b)}{\big( \tau_{\rm sim}(b) \big)^2}
 + \eta^2 =0.
\end{align}
Therefore,
\begin{align}
\tau_{\rm sim}(b) &= \Big| \big[ b \frac{\partial f}{\partial b}/f \big]
\frac{\sqrt{   \tau_{\rm eff}(b)  }}{\eta} \Big| .
\label{tsimb}
\end{align}
We have put the absolute signs $| \cdot |$ 
as the simulation times needs to be
positive.
The same relation holds for $\tau_{\rm sim}(c)$.
\begin{align}
\tau_{\rm sim}(c) &=  \Big| \big[ c \frac{\partial f}{\partial c}/f \big]
\frac{\sqrt{   \tau_{\rm eff}(c)  }}{\eta} \Big|.
\label{tsimc}
\end{align}
We can sum up
Eqs.~(\ref{tsimb}) and (\ref{tsimc}) and use $\tau_{\rm sim}(a)=
\tau_{\rm sim}(b)+\tau_{\rm sim}(c)$ which gives the solution for $\eta$
\begin{align}
\eta=\frac{1}{\tau_{\rm sim}(a)}
\Big( \Big| \big[ b \frac{\partial f}{\partial b}/f \big]
\sqrt{   \tau_{\rm eff}(b)  } \Big|
+  \Big|  \big[ c \frac{\partial f}{\partial c}/f \big]
\sqrt{   \tau_{\rm eff}(c)  } \Big|  \Big).
\label{relq}
\end{align}
Then, substitution of Eq.~(\ref{relq}) in Eqs.~(\ref{tsimb},\ref{tsimc}) 
completes
the equations for $\tau_{\rm sim}(b)$ and $\tau_{\rm sim}(c)$. 
Substitution of these two equations
in Eq.~(\ref{d2a}) results in
\begin{align}
\tau_{\rm eff}(a)=\Big( \Big|
 \big[ b \frac{\partial f}{\partial b}/f \big] \Big| \sqrt{ \tau_{\rm eff}(b)  }
+ \Big|
 \big[ c \frac{\partial f}{\partial c}/f \big] \Big| \sqrt{ \tau_{\rm eff}(c)  }
 \Big)^2.
\label{ect2sim}
\end{align}
The efficiency time $\tau_{\rm eff}(a)$ of Eq~(\ref{ect2sim}) is always strictly less than
or equal to $\tau'_{\rm eff}(a)$ and $\tau''_{\rm eff}(a)$ of
Eqs.~(\ref{teffa1},\ref{teffa2}).
These CPU efficiency times
are straightforwardly
generalized to simulation series of any number.
Suppose that the final desired value $a$ is obtained 
by $a=f(a^{(1)}, a^{(2)}, \ldots, a^{(M)})$, where $a^{(s)}$ refers to the exact value that 
should be produced by simulation $s$ and $M$ is the total number of independent
simulations that are needed to determine $a$.
Then, the CPU efficiency times 
Eqs~(\ref{teffa1}-\ref{ect2sim}) yield 
\begin{align}
\tau_{\rm eff}'(a) &=
 M \Big[ \sum_{s=1}^M
 \big[ a^{(s)} \frac{\partial f}{\partial a^{(s)} }/f \big]^2
 \tau_{\rm eff}^{(s)}
  \Big], \nonumber \\
\tau_{\rm eff}''(a) &=
 \Big( \sum_{s=1}^M \tau_{\rm eff}^{(s)} \Big)
\Big[ \sum_{s=1}^M \big[ a^{(s)} \frac{\partial f}{\partial a^{(s)}}/f \big]^2
 \Big], \nonumber \\
\tau_{\rm eff}(a)&=\Big(  \sum_{s=1}^M \Big|
 \big[ a^{(s)} \frac{\partial f}{\partial a^{(s)}}/f \big] \Big|
\sqrt{ \tau_{\rm eff}^{(s)}  } \Big)^2,
\label{teffa12M}
\end{align}
where $\tau_{\rm eff}^{(s)}=\tau_{\rm eff}(a^{(s)})$ is the efficiency time of simulation $s$. 
For example, in many methods,  the final value $a$ is given  
by a product of simulation results: 
$a=\prod_{s=1}^M a^{(s)}$. 
Then, 
$|\big[ a^{(s)} \frac{\partial f}{\partial a^{(s)}}/f \big]|=1$
for any $s$ after which Eqs.~(\ref{teffa12M}) become
 
\begin{align}
 \tau_{\rm eff}'(a)=\tau_{\rm eff}''(a) &= 
M \Big( \sum_{s=1}^M   \tau_{\rm eff}^{(s)} \Big)
\label{effprod12}
\end{align}
and
\begin{align}
 \tau_{\rm eff}(a) &=  \Big( \sum_{s=1}^M \sqrt{  \tau_{\rm eff}^{(s)}}
\Big)^2.
\label{effprod}
\end{align}
The same Eqs.~(\ref{effprod12},\ref{effprod}) are valid for the case
$a=\prod_{s=1}^J a^{(s)}/\prod_{s'=J+1}^M a^{(s')}$ with any $J \in [0,M]$.

Whenever $\tau_{\rm eff}^{(s)}$ is the same for all $s$, 
Eq.~(\ref{effprod12}) and (\ref{effprod}) become identical and equal to
\begin{align}
\tau_{\rm eff}(a)= \tau_{\rm eff}'(a)=\tau_{\rm eff}''(a) &=
M^2  \tau_{\rm eff}^{(s)}.
\label{effconst}
\end{align}

\section{Efficiency of US techniques}\label{USCPU} 
We will derive the efficiency time $\tau_{\rm eff}^{(s)}$
of a single window in US as depicted in Fig.~\ref{umbrel}. 
The factor that needs to be computed in simulation $s$ is
$a^{(s)}\equiv
\lo w_s \rc_{W_s}/\lo  w_{s-1} \rc_{W_s}=  b/c$. Here,
$b=\lo \hat{b} \rc_{W_s}$, $\hat{b}=w_s$,
$c=\lo \hat{c}\rc_{W_s}$, and $\hat{c}=w_{s-1}$.
Then, we can use
Eq.~(\ref{taucomij}) with $i=1$ and $j=-1$
\begin{align}
\tau_{\rm eff}^{(s)}&= \tau_{\rm eff}(b)+\tau_{\rm eff}(c)-\frac{2}{bc} 
{\rm Cov}(\hat{b},\hat{c}) (1+ n_c^{bc}+n_c^{cb}) \tau_{\rm cyc}
\label{covij}
\end{align}
 and applying the assumptions Eqs.~(\ref{assum},\ref{asstaucyc}) gives
 \begin{align}
\tau_{\rm eff}^{(s)}&= \tau_{\rm eff}(b)+\tau_{\rm eff}(c)-\frac{2}{bc} 
{\rm Cov}(\hat{b},\hat{c})  {\mathcal N_C}.
\label{covij2}
\end{align}
Via  Eq.~(\ref{defcov}), Eq.~(\ref{taubin}),  
and Eqs.~(\ref{assum},\ref{asstaucyc}) we get 
\begin{align}
\tau_{\rm eff}(b)&=\frac{1-b }{b} {\mathcal N_C}, \quad 
\tau_{\rm eff}(c) = \frac{1-c}{c}   {\mathcal N_C},\nonumber \\
{\rm Cov}(\hat{b},\hat{c}) &=
\lo w_{s-1} w_{s} \rc_{W_s} - bc. \label{teffcov}
\end{align}
Then using Eq.~(\ref{windows1}) and~(\ref{V2D})  
and writing $\lambda_R=\lambda^*-\frac{1}{2} W$ and $\alpha \equiv \beta 2 H/W$
 we arrive at 
 \begin{align}
b&=\lo w_s \rc_{W_s}=
\frac{\int_{s \Gamma }^{s \Gamma+\gamma} 
\ud \lambda \, e^{-\alpha \lambda}}{\int_{(s-1) \Gamma }^{s \Gamma+\gamma} 
\ud \lambda \, 
e^{-\alpha \lambda}} = 
\frac{1-e^{-\alpha \gamma }}{e^{\alpha \Gamma }-e^{-\alpha \gamma}}, 
\nonumber \\
c&= \lo w_{s-1} \rc_{W_s}=
\frac{\int_{(s-1) \Gamma }^{(s-1) \Gamma+\gamma} 
\ud \lambda \, e^{-\alpha \lambda }}{
\int_{(s-1) \Gamma }^{s \Gamma+\gamma} \ud \lambda \,
e^{-\alpha\lambda}} = 
 \frac{ e^{\alpha \Gamma } ( 1-e^{-\alpha \gamma })}{e^{\alpha \Gamma }-e^{-\alpha \gamma }}.
 \label{bcUS}
\end{align}
 Moreover, $\lo w_{s-1} w_{s} \rc_{W_s}$
is only nonzero in case $\gamma > \Gamma$.
Hence,
\begin{align}
\lo w_{s-1} w_{s} \rc_{W_s} &= \theta(\gamma-\Gamma)
\frac{\int_{s \Gamma }^{(s-1) \Gamma+\gamma}
\ud \lambda \, e^{-\alpha \lambda}}{\int_{(s-1) \Gamma }^{s \Gamma+\gamma}
\ud \lambda \,
e^{-\alpha \lambda}} 
\nonumber \\
&=\theta(\gamma-\Gamma) \frac{1-e^{-\alpha( \gamma-\Gamma) }}
{e^{\alpha \Gamma }-e^{-\alpha \gamma }}.
\label{partcov} 
\end{align}
Substitution of Eqs~(\ref{bcUS},\ref{partcov}) in Eqs.~(\ref{teffcov}) and, after that, 
substitution of    Eqs.~(\ref{teffcov})  in Eq.~(\ref{covij2}) yields Eq.~(\ref{taueffas}).\\

In order to
derive the efficiency time for an ensemble average  $a=\lo \hat{a}(x) \rc$
when it is obtained by a weighted ensemble via 
$a=\lo \hat{a} \Omega^{-1} \rc_\Omega/ \lo \Omega^{-1} \rc_\Omega$,  
we write again $a=b/c$ with 
$b=\lo \hat{b} \rc_\Omega$, $\hat{b}=\hat{a}\Omega^{-1}$,
$c=\lo \hat{c} \rc_\Omega$, and $\hat{c}=\Omega^{-1}$. 
Applying Eq.~(\ref{wens}) gives $b=\frac{a}{\lo \Omega \rc}$ and 
 $c=\frac{1}{\lo \Omega \rc}$.
Then applying Eqs.~(\ref{std}), (\ref{defcov}) using the ensemble 
$\lo \ldots \rc_\Omega$ gives:
\begin{align}
\sigma_{\hat{b}}^2 &= \lo \hat{b}^2 \rc_\Omega-\lo b \rc_\Omega^2= 
\frac{\lo \hat{a}^2 \Omega^{-1}\rc}{\lo \Omega \rc }
-\Big(\frac{a}{\lo \Omega \rc}\Big)^2, \nonumber \\
\sigma_{\hat{c}}^2 &= \lo \hat{c}^2 \rc_\Omega-\lo c \rc_\Omega^2=
\frac{\lo  \Omega^{-1}\rc}{\lo \Omega \rc }
-\Big(\frac{1}{\lo \Omega \rc}\Big)^2, \nonumber \\
{\rm Cov}(\hat{b},\hat{c})&=\lo \hat{b}\hat{c} \rc_\Omega-
\lo \hat{b} \rc_\Omega\lo \hat{c} \rc_\Omega=\frac{\lo a \Omega^{-1} \rc}{\lo \Omega \rc}
-\frac{ a  }{\lo \Omega \rc^2}. \label{sigmacovw}
\end{align}
Substitution of Eqs.~(\ref{sigmacovw}) in Eq.~(\ref{taueff}) yields
\begin{align}
\tau_{\rm eff}(b)&=\Big( \frac{\lo\hat{a}^2 \Omega^{-1} \rc \lo \Omega \rc}{a^2}-1 \Big) 
\Big[1+2 n_c^{bb}\Big] \tau_{\rm cyc}, \nonumber \\
\tau_{\rm eff}(c)&=\Big( \lo \Omega^{-1} \rc \lo \Omega  \rc-1 \Big) 
\Big[1+2 n_c^{cc}\Big] \tau_{\rm cyc}. \label{teffbcw}
\end{align}
Substitution of Eqs.~(\ref{sigmacovw},\ref{teffbcw}) in Eq.~(\ref{covij}) yields Eq.~(\ref{teffbias}).

To obtain the optimal biasing function, we add 
a Lagrange multiplier to Eq.~(\ref{teffbias2})
to fulfill the normalization constraint
\begin{align}
\tau_{\rm eff}(a) &=\Big\{
\frac{\lo \hat{a}^2 \Omega^{-1}\rc }{\lo a \rc^2}
+ \lo  \Omega^{-1}\rc  - 2  \frac{\lo \hat{a} \Omega^{-1}\rc }{\lo a \rc}
\Big\} {\mathcal N_C}
\nonumber \\
&+ \eta^2 \Big\{ \lo \Omega \rc -1 \Big\}. \label{taueffbiasL}
\end{align}
The functional derivative of $\lo \hat{a} f(\Omega) \rc$ to $\Omega$ for any operator $\hat{a}$ and function $f$ 
is given by
\begin{align}
\frac{\delta \lo a f(\Omega) \rc}{\delta \Omega}= \frac{a(x) f'(\Omega) e^{-\beta E(x)}}{
\int e^{-\beta E(x)} }=a(x) f'(\Omega) P(x)
\end{align}
with $P(x)=\exp(-\beta E(x))/\int {\mathrm d}x \exp(-\beta E(x))$.
Hence taking $\delta \tau_{\rm eff}/\delta \Omega=0$ in Eq.~(\ref{taueffbiasL})
results in 
\begin{align}
\Big\{ \Big[ -\frac{1}{a^2} \frac{\hat{a}^2}{\Omega^2(x)}
- \frac{1}{\Omega^2(x)}+2\frac{1}{a} \frac{\hat{a}(x)}{\Omega^2(x)} \Big] 
{\mathcal N_C}+\eta^2 \Big\}P(x)=0
\end{align}
which has as solution
\begin{align}
\Omega^2(x)=\frac{{\mathcal N_C}}{\eta^2}  \Big[ \frac{\hat{a}^2(x)}{a^2} 
+ 1-2\frac{\hat{a}(x)}{a} \Big]=\frac{{\mathcal N_C}}{\eta^2} \Big( 1 - 
\frac{\hat{a}(x)}{a} \Big)^2. \label{w2}
\end{align}
$\eta^2$ can be obtained, if necessary, via the normalization requirement $\lo \Omega \rc=1$. 
Moreover,  from Eq.~(\ref{w2}) follows directly Eq.~(\ref{optw}) as $\Omega$ should also obey $\Omega(x) > 0$ for each $x$.\\

\section{TIS pathlength}
\label{app_effTIS}
The TIS pathlength~(\ref{GgTIS}) 
can be obtained as follows. Say $v$ is the velocity at the foot of the barrier ($\lambda_0$) at a time 
$t=0$. The 'returning time' is obtained by solving following equation
$\lambda(x_t)=\lambda_0+v t + \frac{F}{2m}t^2=\lambda_0+v t - \frac{H}{m W}t^2=\lambda_0$, which has a
solution for $t=\frac{W m v}{H}\equiv L(v)$. 
Within the path ensemble $s$ all trajectories should cross $\lambda_s$. Hence, at the foot of the barrier the kinetic energy $\frac{1}{2} m v^2$ must be larger than $E \equiv 2H (\lambda_s-\lambda_0)/W$. 
Therefore, for the average pathlength we can write
\begin{align}
\tau_{\rm path} &=\frac{1}{\Delta t} \frac{\int_{\sqrt{2 E/m}}^{\infty} L(v) v e^{-\beta v^2} }
{\int_{\sqrt{2 E/m}}^{\infty}  v e^{-\beta v^2} }=
\frac{W m }{\Delta t H}
\frac{\int_{\sqrt{2 E/m}}^{\infty}  v^2 e^{-\beta \frac{1}{2} m v^2} }
{\int_{\sqrt{2 E/m}}^{\infty}  v e^{-\beta \frac{1}{2} m v^2} } \nonumber \\
&= 
\frac{W m }{\Delta t H}
\frac{2 \sqrt{\beta E }+\sqrt{\pi} e^{\beta E} \,  
{\rm erfc}(\sqrt{\beta E})}{\sqrt{2 \beta m}} \nonumber \\
& \approx \frac{W m }{\Delta t H}  \frac{\sqrt{2} \Big( \sqrt{E}+
\sqrt{\frac{\pi}{\beta (4+\beta E \pi) }}
 \Big)}{\sqrt{m}} \approx  \frac{W}{\Delta t H} \sqrt{2 E m}.
\label{TISpath} 
\end{align}
with $\Delta t$ the MD timestep that is required to express $\tau_{\rm path}$ as an integer representing 
the average number of discrete timesteps. In Eq.~(\ref{TISpath}), we have first used the 
approximation~\cite{mathworld}
\begin{align}
{\rm erfc}(x) \approx \frac{2}{\sqrt{\pi}} \frac{e^{-x^2}}{x+\sqrt{x^2+4/\pi}}
\label{erfc}
\end{align}
and, then, neglected the ${\mathcal O}(E^{-1/2})$ terms.
Using $E = 2H (\lambda_s-\lambda_0)/W$ and $\tau_{\rm path}=G (\lambda_s-\lambda_0)^g$
results in Eq.~(\ref{GgTIS}).

\begin{widetext}
\section{List of symbols} \label{alos}
\begin{tabular}{ll}
$T$ & temperature \\
$\beta=1/T$ & inverse temperature \\
$k_B$ & Boltzmann constant \\
$r$ & configuration point\\
$p$ & momentum point\\ 
$x=(r,p)$ & phase point \\
$m$ & particle mass \\
$V(r)$& potential energy of $r$\\
$E(x)=V(r)+p^2/2m$ & total energy of $x$ \\
$\rho(x)$ & equilibrium distribution, $\rho(x) = e^{-\beta E(x)}$ 
for Boltzmann statistics\\
$x_t$ &  phase point at time $t$\\
$\Delta t$ & MD timestep \\
$X$ &path consisting of discrete 
timeslices: $ \{x_{-\tau^b \Delta t},\ldots;x_0; 
\ldots, x_{\tau^f \Delta t}\}$ \\
$\tau^b[X],\tau^f[X]$ & the start and end time index  of path $X$\\
$p_{\rm n}(x \rightarrow y)$ & the probability density to go to $y$ from $x$
in one timestep by MD\\
$p_{\rm n}(x \leftarrow y)$ & the chance that when you are in $x$, you came from $y$ one timestep before\\
$p_{\rm s}(x \rightarrow y), p_{\rm s}(x \leftarrow y)$ & 
same hopping rates for two consecutive simulation cycles using MD/MC  \\
$P[X]$ & weight of the path $X$ \\
$q$ & point in either phase or path space\\
$q_i$ & phase/path point generated after the $i$-th simulation cycle\\
$\lo \ldots \rc $ & ensemble average  \\
$\lo \ldots \rc_\Omega $ 
& weighted ensemble average using weight function $\Omega(x)$  \\
$a$ & exact value of an observable\\
$\hat{a}$(q) &  generic operator that determines $a$\\
$a_i=\hat{a}(q_i)$ & function value of the $i$-th simulation cycle\\
$\bar{a}(n)$ & average function value over $n$ simulation cycles \\
$\Delta a(n)$ & absolute error in $a$ after $n$ cycles\\
$\delta a(n)$ & relative error in $a$ after $n$ cycles\\
$\sigma_a$ & standard deviation of the distribution $\{ a_i \}$ \\
$\sigma_{\bar{a}(n)}$ & standard deviation of the distribution 
$\{ \bar{a}(n) \}$ \\
${\rm Var}(\hat{a})$ & variance of $\hat{a}$ \\
${\rm Cov}(\hat{a},\hat{b})$ & covariance of $\hat{a}$ and $\hat{b}$ \\
$C_{a,b}(l)$ & correlation function \\
$n^{a,b}$ & correlation number \\
${\mathcal N_C}$ & effective total correlation \\
$s$ & index of a simulation in a simulation series \\
$a^{(s)}$ & exact value of an observable obtained from simulation $s$\\
$\tau_{\rm cyc}$ & average duration of a simulation cycle \\
$\tau_{\rm sim}$ & total simulation time \\
$\tau_{\rm path}$ & average path length in a path sampling simulation \\
$\tau_{\rm esc}$ & maximum time needed to leave the barrier region\\
$\tau_{\rm eff}(a)$ & lowest computational cost needed to determine 
$a$ with a relative error equal to one\\
$\xi$ & average ratio $\tau_{\rm cyc}/\tau_{\rm path}$\\
$k$ & reaction rate \\
${\mathcal R}$ & unnormalized transmission coefficient \\
$\tilde{k}, \tilde{\mathcal R}$ & time-dependent rate and transmission functions\\
$\chi[X]$ & recrossing correction functional\\
$\lambda(x)$ & reaction coordinate\\
$P(\lambda')$  & probability density to be on the
surface $\{x | \lambda(x)=\lambda'\}$ \\
$P_A(\lambda')$  & probability density to be on the 
surface $\{x | \lambda(x)=\lambda'\}$ given you are in $A$\\
\end{tabular}
\end{widetext}
\begin{widetext}
\begin{tabular}{ll}
$F(\lambda)=-\ln( P(\lambda) )/\beta$ & 
free energy profile along $\lambda$\\
$\lambda^*$ & transition state  value or the maximum in $F(\lambda)$ \\
$\lambda_s$ & value defining interface $s$: $\{x| \lambda(x)=\lambda_s \}$\\
$\lambda_A=\lambda_0$ & interface defining stable state $A$\\
$\lambda_B=\lambda_M$ & interface defining stable state $B$\\
$M$ & total number of simulations used in a simulation series\\ 
${\mathcal P}_A(\lambda|\lambda')$ & crossing probability from interface $\lambda'$ to $\lambda$\\
$R_x, R_y$ & dimensions of the reactant well \\
$W$ & width of the barrier \\
$H$ & height of the barrier \\
$\lambda_x$ & assumed reaction coordinate in the 2D system \\
$\lambda_y$ & important other degrees of freedom  in the 2D system\\
$\lambda_\perp$ & unknown ideal reaction coordinate \\
$\theta$ & 
angle between $\lambda_x$ and  $\lambda_\perp$ giving the deviation 
from the optimal RC\\
$\Gamma, \gamma$ & dimensions of rectangular windows used in US\\
$w_s(x), W_s(x)$ & block functions defined by Eq.~(\ref{windows1})  \\
$P_{\lambda_s}^{\rm TI/US}(\lambda_y)$ & sampling distribution along
$\lambda_y$ at the surface $\lambda_s$ when using TI or US\\
$P_{\lambda_s}^{\rm TIS}(\lambda_y)$ & sampling distribution 
along $\lambda_y$
of first crossing points with surface $\lambda_s$
when using TIS \\
$g,G$ & exponent and pre-exponential factor the assumed behavior of $\tau_{\rm path}^{(s)}$ \\
$h_{\mathcal A}(x)$ & history dependent function that measures whether $x$ 
was more recently in $A$ than in $B$\\
$h_{i,j}^b(x)$ &
history dependent function that measures whether $x$
has crossed $\lambda_i$ more recent than $\lambda_j$\\
$h_{i,j}^f(x)$ &
future dependent function that measures whether $x$
will cross $\lambda_i$ before $\lambda_j$\\
$\phi_A$ & flux function through $\lambda_A$ equal to 
$\delta(\lambda(x)-\lambda_A) \dot{\lambda} \theta( \dot{\lambda} )$
\end{tabular}
\end{widetext}
\section{List of abbreviations}\label{aloa}
\begin{tabular}{ll}
BC & Bennett-Chandler formalism\\
BC2 & history dependent BC \\
EPF & the effective positive flux \\
FFS & forward flux sampling \\
MD & molecular dynamics \\
MC & Monte Carlo \\
RF & reactive flux method \\
RC & reaction coordinate \\
TI & thermodynamic integration \\
TIS & transition interface sampling \\ 
TPS & transition path sampling \\
TS & transition state \\
TST & transition state theory \\
US & umbrella sampling
\end{tabular}\\
\null \\ \null \\ \null \\ \null \\  \null \\
\null \\ \null \\ \null \\ \null \\  \null \\
\bibliographystyle{prsty}

\end{document}